\documentclass[preprint,prd,amsmath,amssymb,aps,nofootinbib]{revtex4-1}

\usepackage{graphicx}
\usepackage{color}
\usepackage{bm}
\usepackage{natbib}
\usepackage{enumerate}
\def\aap{Astron.\ Astrophys.\ }
\def\aapr{Astron.\ Astrophys.\ Rev.\ }
\def\apj{Astrophys.\ J.\ }
\def\apjl{Astrophys.\ J.\ Lett.\ }
\def\apjs{Astrophys.\ J.\ Supp.\ }
\def\aj{Astron.\ J.\ }
\def\mnras{Mon.\ Not.\ Roy.\ Astron.\ Soc.\ }

\def\prd{Phys.\ Rev.\ D\ }
\def\prl{Phys.\ Rev.\ Lett.\ }

\def\araa{Annu.\ Rev.\ Astron.\ Astrophys.\ }
\def\jcap{J.\ Cosmol.\ Astropart.\ Phys.\ }
\def\ssr{Space\ Sci.\ Rev.\ }

\begin{document}
\title{Explanation of the knee-like feature in the DAMPE cosmic $e^-+e^+$ energy spectrum}

\author{Kun Fang}
\affiliation{Key Laboratory of Particle Astrophysics, Institute of High Energy Physics, Chinese Academy of Sciences, Beijing 100049, China}
\affiliation{School of Physics, University of Chinese Academy of Sciences, Beijing 100049, China}

\author{Xiao-Jun Bi}
\affiliation{Key Laboratory of Particle Astrophysics, Institute of High Energy Physics, Chinese Academy of Sciences, Beijing 100049, China}
\affiliation{School of Physics, University of Chinese Academy of Sciences, Beijing 100049, China}

\author{Peng-Fei Yin}
\affiliation{Key Laboratory of Particle Astrophysics, Institute of High Energy Physics, Chinese Academy of Sciences, Beijing 100049, China}

\date{\today}

\begin{abstract}
The DArk Matter Particle Explorer (DAMPE), a space-based high precision
cosmic ray detector, has just reported the new measurement of the total
electron plus positron energy spectrum up to 4.6 TeV. A notable feature
in the spectrum is the spectral break at $\sim 0.9$ TeV, with the
spectral index softening from $-3.1$ to $-3.9$. Such a feature is very
similar to the knee at the cosmic nuclei energy spectrum. In this work
we propose that the knee-like feature can be explained naturally
by assuming that the electrons are accelerated at the supernova remnants (SNRs) and released when the SNRs die out with lifetime around $10^5$ years.
The cut-off energy of those electrons have already decreased to several TeV
due to radiative cooling, which may induce the observed TeV spectral break.
Another possibility is that the break is induced by a single nearby old SNR.
Such a scenario may bring a large electron flux anisotropy that is
observable by the future detectors. We also show that a minor
part of electrons escaping during the acceleration in young and nearby
SNRs are able to contribute to several TeV or higher energy region of the spectrum.
\end{abstract}

\maketitle

\section{Introduction}
\label{sec:intro}
The DArk Matter Particle Explorer (DAMPE), which is one of the new generation
space-borne instruments for measuring cosmic rays (CRs), has been running in
orbit for nearly two years. Recently, the DAMPE Collaboration published their
first result of the high energy CR electron plus positron ($e^-+e^+$) spectrum from 25 GeV to 4.6 TeV, with
the data recorded between December 27, 2015 and June 8, 2017 \cite{nature}. The measurement of DAMPE is very precise due to
its unprecedentedly high energy resolution and strong electron/proton discrimination \cite{dampe}. This measurement provides the direct evidence of a spectral break at $\sim 0.9$ TeV for the first time,
which is consistent with the spectral break around 1 TeV detected indirectly by
ground-based experiments, such as H.E.S.S and VERITAS \cite{hess09,veritas}.
The DAMPE spectrum shows another interesting feature at $\sim1.4$ TeV, although
it is not significant right now. Meanwhile, the uncertainties above 3 TeV are large at present, and spectral features are still possible at the high energy end.

The most notable and clear feature of the DAMPE spectrum is the break at $\sim 0.9$ TeV. This feature is very similar to the 'knee' of the cosmic nuclei energy spectrum, which has been observed by ground-based experiments for more than 50 years \cite{Amenomori:2011zza,Bartoli:2015vca}. The origin of the knee
is still under debate due to the lack of direct measurement in space with
clear identification of nuclei species. One naturally explain is that the knee is caused by the acceleration limit in the cosmic ray sources \cite{2004APh....21..241H}. Similarly, the break feature at the electron spectrum may also be caused by the cut-off energy of electrons, which is much lower than that of nuclei because of the large radiative energy loss of the electron.

Supernova remnants (SNRs) have long been considered as the major source of
galactic CRs \cite{1934PNAS...20..259B,1961PThPS..20....1G,2013A&ARv..21...70B}.
The electron injection of an SNR is often assumed to be happened at the birth time of the SNR \cite{dela10,mauro14}. However, this is true only when the delay time (the time from birth to CR injection) of the injection is far less than the age of the SNR. In this work, we consider a more realistic injection picture for SNRs. CRs are largely released from an SNR when the SNR dies as a CR factory at an age of $\sim10^5$ yr \cite{2010APh....33..160C}. Meanwhile, the maximum energy of electrons in a mature SNR should decrease with time, due to significant radiative cooling. As we will show, an evolution time of $\sim10^5$ yr corresponds to a cut-off energy of $\sim$ TeV for electrons, which can explain the TeV break detected by DAMPE.

On the other hand, spectral features in the electron spectrum are often expected in the TeV region, which could be produced by young and nearby CR sources \cite{shen70,koba04}. The final released population mentioned above can hardly account for remarkable spectral features at such high energy.
However, if a minor part of electrons can escape from the SNRs during the acceleration process, the young and nearby SNRs may contribute such a high energy component. We will show below that the spectral index of this population of electrons can be very hard, which is necessary for a distinctive spectral structure in high energies.

In this work, we use the CR electrons released at the final stage of SNRs to explain the DAMPE result, and also consider the contribution from the CR electrons released during the acceleration process. Besides, CR positrons contribute $\sim10\%$ of the total $e^-+e^+$ flux, which have been measured by space-borne detectors with the magnetic spectrometer, such as PAMELA and AMS-02 \cite{2013PhRvL.111h1102A,amsfrac}. The `positron excess' in cosmic rays is one of the most important findings in CR researches. Thus we attempt to explain the DAMPE $e^-+e^+$ spectrum and the positron fraction ($e^+/(e^-+e^+)$) of AMS-02 \cite{amsfrac} simultaneously to obtain a complete picture. We consider pulsar wind nebulae (PWNe) as the sources of high energy positrons. Note that the $e^\pm$ injection spectrum of a PWN could have a rapid drop due to the cooling effect. This mechanism may account for the tentative feature at 1.4 TeV of the DAMPE spectrum, if this feature is actually edge-like.

This paper is organized as follows. In Sec. \ref{sec:prop}, we introduce the
propagation of CR electrons and positrons ($e^{\pm}$) after their injection
into the interstellar medium (ISM), and show the calculation of the $e^\pm$
spectrum at the Earth. In Sec. \ref{sec:source}, we describe the
picture of electron escape for SNRs, and also introduce the positron sources. In Sec. \ref{sec:results}, the fitting results are presented and analyzed. Then we discuss a possible scenario to reproduce the TeV spectral break by a single nearby source in Sec. \ref{sec:mr}. Finally, we conclude in Sec. \ref{sec:conclusion}.

\section{Propagation of Galactic Electrons and Positrons}
\label{sec:prop}
The propagation process of $e^\pm$ can be described by the diffusion equation
with consideration of the radiative cooling during their journey:
\begin{equation}
 \frac{\partial N}{\partial t} - \nabla(D\nabla N) - \frac{\partial}{\partial
E}(bN) = Q \,,
 \label{eq:diff}
\end{equation}
where $N$ is the differential number density of $e^\pm$, $D$ denotes the
diffusion coefficient, $b$ is the energy-loss rate, and the source function of
$e^\pm$ is denoted by $Q$. The data of DAMPE is above 20 GeV, where convection
or reacceleration has little effect on $e^-+e^+$ spectrum \cite{dela09}, so the
relevant terms are not included in Eq. (\ref{eq:diff}). The propagation zone is
set as a cylindrical slab, with a radius of 20 kpc and a half thickness of
$z_h$. The diffusion coefficient is usually assumed to be $D(E)=\beta D_0{(R/\rm
1\,GV)}^{\delta}$, where $D_0$ and $\delta$ are both constants, $\beta$ is the
velocity of particles in the unit of light speed, and $R$ is the rigidity of
$e^\pm$. We adopt the propagation parameters given by the DR2 model (the
revised diffusion reacceleration model) in Ref. \cite{2017PhRvD..95h3007Y}:
$D_0=(2.08\pm0.28)\times10^{28}$ cm$^2$ s$^{-1}$, $\delta=0.500\pm0.012$, and
$z_h=5.02\pm0.86$ kpc. This DR2 model is the best performing model in Ref.
\cite{2017PhRvD..95h3007Y} when fitting to the latest B/C data of AMS-02
\cite{2016PhRvL.117w1102A}. The energy-loss rate has the form of
$b(E)=b_0(E)E^2$, where $b_0(E)$ is decided by synchrotron and inverse Compton
radiation of $e^\pm$. We set the interstellar magnetic field in the Galaxy to be
$1~\mu$G to get the synchrotron term \cite{1994A&A...288..759H, dela10}. The
inverse Compton term refers to the calculation of \cite{schli10}, in which a
relativistic correction to the scattering cross-section is considered.

The general solution of Eq. (\ref{eq:diff}) can be obtained by the method of
Green's function \cite{1964ocr..book.....G}, which is written as
\begin{equation}
 N(E, \bm{r}, t)=\iiint dE_0d\bm{r}_0dt_0\,Q(E_0, \bm{r}_0, t_0)G(E, \bm{r}, t;
E_0, \bm{r}_0, t_0)\,.
\label{eq:N_gen}
\end{equation}
The infinite three-dimensional solution can be a good approximation for high
energy case \cite{koba04,2017ApJ...836..172F}, which takes the form of
\begin{equation}
 G(E, \bm{r}, t; E_0, \bm{r}_0, t_0) = \frac{1}{(\pi\lambda^2)^{3/2}b(E)}{\rm
exp}\left[-\frac{(\bm{r}-\bm{r}_0)^2}{\lambda^2}\right]\delta(t-t_0-\tau)\,,
\end{equation}
where
\begin{equation}
 \tau=\int_{E}^{E_0}\frac{dE'}{b(E')}\simeq\frac{1}{b_0}\left(\frac{1}{E}-\frac{1}{E_0}\right)
 \label{eq:tau}
\end{equation}
and
\begin{equation}
\lambda^2=\int_{E}^{E_0}\frac{D(E')}{b(E')}dE'\simeq\frac{D_0\left[1-(E/E_0)^{
1-\delta}\right]}{b_0(1-\delta)E^{1-\delta}}\,.
 \label{eq:lambda}
\end{equation}
For a point source with burst-like injection, the source function is $Q(E,
\bm{r}, t) = Q(E)\delta(\bm{r}-\bm{r}_s)\delta(t-t_s)$. Setting the location of
the Earth as $r=0$ and the present time as $t=0$, and substituting this source
function expression into Eq. (\ref{eq:N_gen}), we obtain the spectrum
contributed by a source with distance $r$ and age $t$:
\begin{equation}
I(E, r, t)=\frac{c}{4\pi}N(E,t,r)
 =\frac{c}{4\pi^{5/2}\lambda^3}\,\frac{b(E^\star)}{
b(E)}\,{ \rm
exp}\left(-\frac{r^2}{\lambda^2}\right)\,Q(E^\star)\,,
\label{eq:I_s}
\end{equation}
where $E^\star=E/(1-b_0Et)$. As to a point source with
continuous particle injection, the source function is written as $Q(E, \bm{r},
t) = Q(E, t)\delta(\bm{r}-\bm{r_s})$. Similarly, we get the spectrum in this
case:
\begin{equation}
 I(E, r, t) =
\int_{t_{\rm
min}}^{t_{\rm max}}\frac{c}{4\pi^{5/2}\lambda^{3}}\,\frac{b(E^\star)}{b(E)}{\rm
exp}\left(-\frac{r^2}{\lambda^2}\right)\,Q(E^\star,
t_0)dt_0\,,
\label{eq:I_c}
\end{equation}
where $t_{\rm max}$ is the smaller one between the age $t$ and $1/(b_0E)$, and
we set $t_{\rm min}$ to be $r/c$ since Eq. (\ref{eq:diff}) does not
reject superluminal case.

\section{Galactic Sources of Electrons and Positrons}
\label{sec:source}

\subsection{An Electron Escaping Scenario of SNRs}
\label{subsec:snr}
Particles can be boosted to very high energies via the
first-order Fermi acceleration in the shock wave of an SNR
\cite{1977DoSSR.234.1306K,1977ICRC...11..132A,1978MNRAS.182..147B,
1978MNRAS.182..443B,1978ApJ...221L..29B}. Accelerated CR
particles are trapped in the SNR for $\sim 10^5$ years until the shock
dissolves into the ISM. However, adiabatic energy losses during the expansion
of the SNR prevents the released CRs to reach the high energies as
observed, thus the additional particle escape process during the acceleration is required \cite{2010APh....33..307C}. Hence CR particles injected from SNRs can be divided into two populations: CRs escaping from the upstream region during the Sedov phase, and CRs released in the final stage of SNRs
\cite{2010APh....33..160C,2015MNRAS.447.2224B}. In this work, we adopt the
above picture for the electron injection of SNRs to explain the DAMPE data.

\subsubsection{Electron Spectra of SNRs}
To obtain the injection spectrum, we apply the result of Ref.
\cite{2007A&A...465..695Z}, in which the method of asymptotic solution is used
to get analytical expressions for electron spectra in SNRs. Under the
assumption that the energy losses of electrons are dominated by the synchrotron
radiation, the electron momentum distribution in the very high energy range at
the shock can be solved analytically from the transport equation as
\begin{equation}
 f_0(p) = \sqrt{b(p)/p}\,\,{\rm
exp}\left[-\frac{\gamma^2}{u_1^2}\,\left(1+\sqrt{\kappa}\right)^2\,\int_{0}^{p}
\frac{b(p')D(p')}{p'^2} dp'\right]\,,
 \label{eq:f0_gen}
\end{equation}
where $b(p)$ is the downstream synchrotron energy loss rate, $D(p)$ is the
diffusion coefficient in the downstream, $\gamma=3\sigma/(\sigma-1)$ with
$\sigma$ denoting the shock compression ratio, $u_1$ denotes the upstream shock
speed, and $\kappa$ is the ratio of the magnetic field of the upstream to
that of the downstream. In the case of nominal Bohm diffusion with
$D(p)=pc^2/3eB$, where $e$ is the electric charge and $B$ is the downstream
magnetic field, it can be inferred from Eq. (\ref{eq:f0_gen}) that
\begin{equation}
 f_0(p)\propto p^{1/2}\,{\rm exp}(-p^2/p_c^2)\,,
 \label{eq:f0_hi}
\end{equation}
where the cut-off momentum is written as
\begin{equation}
 p_c = 4.3\,{\rm TeV/c}\,\left(\frac{u_1}{300\,{\rm
km\,s}^{-1}}\right)\,\left(\frac{B}{30\,\mu {\rm G}}\right)^{-1/2}\,,
 \label{eq:p_c}
\end{equation}
if we take the standard compression factor $\sigma=4$ and $\kappa=1/\sqrt{11}$
as in Ref. \cite{2007A&A...465..695Z}. In the low energy region, the energy
losses can be neglected, and the transport equation reduces to the simplest
diffusion-convection equation. Then $f_0(p)$ can be expressed by the well-known
form $f_0(p)\propto p^{-4}$ for $\sigma=4$.

In the high energy cut-off region, both the spatially integrated spectrua in the
upstream $F_1(p)$ and downstream $F_2(p)$ have the relations with $f_0(p)$ as
$\propto p^{-1}f_0(p)$, and subsequently $\propto p^{-1/2}\,{\rm
exp}(-p^2/p_c^2)$ \cite{2007A&A...465..695Z}.
While in the low energy part, the momentum distribution of
the upstream is $f_1(x,p)=f_0(p)\,{\rm exp}(u_1x/D_1)$ ($x=0$ for the shock
front and $x<0$ for the upstream region), which means $F_1(p)\propto p^{-3}$
under the Bohm diffusion. Then the downstream integrated
spectrum is approximately given by $F_2(p)\propto p^{-5}$, which can be derived from the boundary condition at the shock front \cite{2007A&A...465..695Z}. The integrated spectra change gradually from the single power-law form to the form of $p^{-1/2}\,{\rm exp}(-p^2/p_c^2)$ when
$p$ tends to be $p_c$. Since in the cut-off region the exponential term
overwhelms the $p^{-1/2}$ term, we can approximate the integrated spectra in the
whole energy range by
\begin{equation}
 F_1(p)\propto p^{-3}\,{\rm exp}(-p^2/p_c^2)\,,
 \label{eq:F1}
\end{equation}
and
\begin{equation}
 F_2(p)\propto p^{-5}\,{\rm exp}(-p^2/p_c^2)\,.
 \label{eq:F2}
\end{equation}

\subsubsection{Electrons Released at the Final Stage}
\label{subsubsec:pop_A}
As we have described above, electrons injected by SNRs may consist of a final
released component (population A hereafter) and a component escaping during the
acceleration (population B hereafter). We assume all the SNRs dissolve into the ISM and massively release their electrons in a same evolution time of $t_{\rm end}$ (the typical age of Sedov phase of an SNR is $\sim 5\times10^4$ yr
\cite{yamazaki06}, which means $t_{\rm end}\gtrsim 5\times10^4$ yr). This
indicates that the population A can only be contributed by SNRs older than $t_{\rm end}$, and thus the injection time of a source with an age of $t$ is $t_0=t-t_{\rm end}$.
We use the smooth radial distribution of Ref. \cite{l04} for all the SNRs with
$t>t_{\rm end}$ to calculate the spectrum of the population A.

For smooth distributed sources with the continuous injection, the observed
spectrum can also be derived by Eq. (\ref{eq:N_gen}):
\begin{equation}
 I(E)=\int_{0}^{\frac{1}{b_0E}}dt_0\int_{0}^{r_{\rm
max}}dr_0\int_{0}^{2\pi}d\varphi_0 \,
\frac{c}{4\pi^{5/2}\lambda^{3}}\,\frac{b(E^\star)}{b(E)}{\rm
exp}\left(-\frac{r_0^2}{\lambda^2}\right)\,f\,\rho(r_0,\varphi_0)\,r_0\,Q(E^\star)\,,
 \label{eq:I_sc}
\end{equation}
where $f=4$ century$^{-1}$ galaxy$^{-1}$ is the explosion rate of SNR in the Galaxy,
$\rho(r,\varphi)$ is the normalized distribution centered
on the solar system, $E^\star=E/(1-b_0Et_0)$, and $r_{\rm max}=$ min\{20
kpc, $ct_0$\}. The population A should be a mixture of upstream and
downstream electrons, so according to Eq. (\ref{eq:F1}) and Eq. (\ref{eq:F2}),
we write the injection spectrum of the population A as
\begin{equation}
 Q(E) = Q_{0, {\rm A}}(E/{\rm 1\,GeV})^{-\gamma_{\rm A}}{\rm exp}(-E^2/E_c^2)\,,
 \label{eq:Q_a}
\end{equation}
where $Q_{0, {\rm A}}$ is the normalization, $\gamma_{\rm A}$ is the power-law
index, and $E_c$ denotes the cut-off energy which is decided by Eq.
(\ref{eq:p_c}). For the integrated spectra of the upstream and downstream, we have $\gamma_{\rm
A}=1$ and $\gamma_{\rm A}=3$ respectively. However, the
population A should be dominated by the
downstream electrons, so $\gamma_{\rm A}$ should be between 2 and 3.

For an SNR which has gone through the free expansion phase, the shock velocity
can be expressed as
\begin{equation}
 u_1=3.9\times10^8\,{\rm
cm\,s^{-1}}\,E_{51}^{1/3}n_{\rm ISM}^{-1/3}\left(\frac{t_{\rm
evo}}{10^3\,{\rm
yr}}\right)^{-3/5}\,
 \label{eq:u1_2}
\end{equation}
for $t_{\rm evo}<4\times10^4E_{51}^{4/17}n_{\rm ISM}^{-9/17}$ yr, and
\begin{equation}
 u_1=2.3\times10^7\,{\rm
cm\,s^{-1}}\,E_{51}^{11/51}n_{\rm ISM}^{-4/17}\left(\frac{t_{\rm
evo}}{10^5\,{\rm
yr}}\right)^{-2/3}\,
 \label{eq:u1_3}
\end{equation}
for $t_{\rm evo}>4\times10^4E_{51}^{4/17}n_{\rm ISM}^{-9/17}$ yr,
where $t_{\rm evo}$ is the evolution time of the SNR, $E_{51}$ is the initial
energy of SN ejecta in units of $10^{51}$ erg, and
$n_{\rm ISM}$ is the number density of ISM in units of 1 cm$^{-3}$
\cite{yamazaki06}. Here we take the typical value $E_{51}=n_{\rm ISM}=1$ for
them. If we assume a typical magnetic field of 10 $\mu$G for the SNR
environment, then we have the magnetic field in the downstream as
$B=30$ $\mu$G for $\kappa=1/\sqrt{11}$. Then for an
SNR stepping into the radiative phase, the cut-off energy $E_c$ is simply determined
by $t_{\rm evo}$ through Eq. (\ref{eq:p_c}) and Eq. (\ref{eq:u1_3}).
Taking the typical lifetime of SNRs as $t_{\rm end}=10^5$ yr,
the corresponding $E_c(t_{\rm end})$ is 3.2 TeV,
and the break in the observed spectrum should begin at lower energy due to cooling of electrons in the propagation. Such a behavior is consistent with the TeV break in the DAMPE data. In the following fitting procedure, we set $Q_{0, {\rm A}}$, $\gamma_{\rm A}$, and $t_{\rm end}$ as the free parameters of the population A.

\begin{table}[t]
\centering
 \begin{tabular}{cccc}
  \hline
  Name & Other Name & $r$(kpc) & $t$(kyr) \\
  \hline
  G65.3+5.7 &  - & 0.9 & 26 \\
  \hline
  G74.0-8.5 & Cygnus Loop & 0.54 & 10 \\
  \hline
  G114.3+0.3 &  - & 0.7 & 7.7\\
  \hline
  G127.1+0.5 & R5 & 1.00 & 25\\
  \hline
  G156.2+5.7 & - & 1.00 & 20.5\\
  \hline
  G160.0+2.6 & HB9 & 0.8 & 5.5\\
  \hline
  G263.9-3.3 & Vela & 0.29 & 11.3\\
  \hline
  G266.2-1.2 & Vela Jr. & 0.75 & 3\\
  \hline
  G347.3-0.5 & RX J1713.7-3946 & 1.00 & 3.2\\
  \hline
 \end{tabular}
 \caption{The name,  distance, and age of SNRs within 1 kpc included by the
Green's catalog. One can refer to Ref. \cite{mauro14} and references therein
for parameters of these sources. For parameters given in the form of interval
in Ref. \cite{mauro14}, we take the mean values here.}
\label{tab:SNRs}
\end{table}

\subsubsection{Electrons Escaping during the Acceleration}
\label{subsubsec:pop_B}
The population B are accelerated electrons that are not advected to the
downstream, but escape from the upstream of the shock of an SNR.
Obviously, the injection of the population B should be a continuous process
for each SNR. For those SNRs with age $t>t_{\rm end}$, the injections last
from $t$ to $t-t_{\rm end}$, while for a source younger than $t_{\rm end}$,
the injection should last from $t$ to now. The arrival spectrum of an
SNR with continuous injection can be calculate with Eq. (\ref{eq:I_c}). For
$t>t_{\rm end}$, $t_{\rm max}={\rm min}\{t, 1/b_0E\}$ and $t_{\rm min}={\rm
max}\{t-t_{\rm end},r/c\}$; while for $t<t_{\rm end}$, $t_{\rm max}$ and
$t_{\rm min}$ are defined as in Sec. \ref{sec:prop}. The injection spectrum can
be expressed as
\begin{equation}
 Q(E,t_0)=\dot{Q}_{0, {\rm B}}(E/{\rm 1\,GeV})^{-1}{\rm
exp}\{-E^2/[E_c(t-t_0)]^2\}\,,
 \label{eq:Q_b}
\end{equation}
where $t_0$ is the injection time, $\dot{Q}_{0, {\rm B}}$ is the
constant rate of injection, which has the unit of GeV$^{-1}$ s$^{-1}$.

Eq. (\ref{eq:F1}) indicates that the integrated spectrum in the upstream takes the form of $\propto E^{-1}$ in the low energy part. We adopt this hard spectrum as the power-law term of the injection spectrum of the population B, as shown in Eq. (\ref{eq:Q_b}). We have mentioned in Sec. \ref{subsubsec:pop_A} that the
momentum distribution of the upstream takes the form of $f_1(x,p)=f_0(p)\,{\rm
exp}[u_1x/D_1(p)]$ in the low energy region, which implies a harder spectrum for
a farther distance from the shock front. This is because the diffusion scale is
larger for electrons with higher energy. If one believes a larger diffusion
scale equals to a larger chance of escape, the injection spectrum should be even
harder than $\propto E^{-1}$. We do not intend to investigate the specific
mechanisms of particle escape in the present work, but discuss a scenario of
free escape in Appendix \ref{app:escape}.

The cut-off energy for a continuous injection SNR is time dependent, which
can be inferred from Eq. (\ref{eq:p_c}), Eq. (\ref{eq:u1_2}), and Eq.
(\ref{eq:u1_3}). The assumptions about $E_0$, $n_{\rm ISM}$, and $B$ are kept
the same with those in Sec. \ref{subsubsec:pop_A}. For
an SNR with an age of $t$, when it releases electrons at $t_0$, it
has evolved for a time of $t_{\rm evo}=t-t_0$. Thus the cut-off energy should be
$E_c(t-t_0)$.

We treat $\dot{Q}_{0, {\rm B}}$ as a free parameter. For old sources contributing to the
population B, their contributions are submerged in the
population A. This means that their $\dot{Q}_{0,{\rm B}}$ cannot be restrict by the
data. On the other hand, the high energy end of DAMPE still reserves
the possibility of spectral features above $\sim3$ TeV. Such features can be
produced by young and nearby sources contributing to the population B, so we only focus on this kind of SNRs. SNRs within 1 kpc included by the Green's catalog \cite{green} are displayed in Table \ref{tab:SNRs}. All of them are younger than
$\sim5\times10^4$ yr. We choose them as the sources of the population B, and assume that they share a common $\dot{Q}_{0,{\rm B}}$ in the following fitting procedure.

\subsection{Positrons}
\label{subsec:positron}
Although SNRs are expected to mainly account for the features of the $e^-+e^+$
spectrum of DAMPE, extra positron sources are needed to explain the
'positron excess' observed by PAMELA and AMS-02. In this
work, we use a single powerful pulsar as the high energy $e^\pm$ source, and
test two different known pulsars. Besides, the calculation of the secondary
positrons is also briefly mentioned.

\subsubsection{Pulsars}
\label{subsubsec:psr}
Electron/positron pairs are produced in the strong magnetic field of pulsars
through the electromagnetic cascade \cite{1982ApJ...252..337D}. The
details of the injection spectrum of pulsars is described in Appendix
\ref{app:pulsar}. In this work, the injection spectrum of a pulsar can be
determined by three parameters: the conversion efficiency from the total
spin-down energy of the pulsar to injected electrons \textit{or} positrons,
$\eta_{\rm pwn}$, the spectral index of its PWN, $\gamma_{\rm pwn}$, and the cut-off energy of its PWN, $E_{c, {\rm pwn}}$.

Firstly, we adopt PSR J0940-5428 to be the single source of the
positron excess. The reason of choosing this pulsar is given in Appendix
\ref{app:pulsar}. PSR J0940-5428 is
a Vela-like pulsar \cite{2007AJ....134.1231C} with a distance
of 0.38 kpc, an age of $4.2\times10^4$ yr, and a spin-down energy of
$1.34\times10^{49}$ erg. It is also detected as a
$\gamma$-ray pulsar \cite{2011arXiv1110.1210H}. There is no further information
about its injection spectrum provided by observations. So in the following
model discussions, we set all the parameters of the injection spectrum of PSR
J0940-5428, i.e., $\eta_{\rm pwn}$, $\gamma_{\rm pwn}$, and $E_{\rm c, pwn}$ as
free variables. As we assume a continuous injection for pulsars, the spectrum
at the Earth should be calculated by Eq. (\ref{eq:I_c}).

Geminga pulsar is expected to be a significant $e^\pm$ contributor
\cite{2009PhRvL.103e1101Y,2013ApJ...772...18L,2017arXiv170208436H}, which
is nearby ($r=0.25$ kpc) and relatively old ($t=3.42\times10^5$ yr) with a large enough spin-down energy of $1.23\times10^{49}$ erg. We consider
Geminga as an alternative single positron source, and set its conversion efficiency
$\eta_{\rm pwn}$ and power-law index $\gamma_{\rm pwn}$ as free parameters. The
cut-off energy $E_c$ is fixed to be 50 TeV, which is consistent with the observation of HAWC
along with Milagro \cite{2009ApJ...700L.127A,2017ApJ...843...40A}.

Recently, there are arresting discussions about Geminga, focusing on whether its injected
$e^\pm$ can really reach the Earth \cite{Abeysekara911,2017arXiv171107482H}. The
spatial profile of $\gamma$-ray around Geminga implies a much slower diffusion
of $e^\pm$ in this region than that in the ISM \cite{Abeysekara911}. However, if
this slow-diffuse region is fairly small, Geminga may still contribute
considerably to the observed $e^-+e^+$ spectrum \cite{2017arXiv171107482H}.

\subsubsection{Secondary Positrons}
\label{subsubsec:second}
Secondary positrons produced by the inelastic collision between CR nuclei
and ISM
are treated as the background of the exotic positron spectrum. We
use the method of \cite{dela09} to obtain the spectrum of secondary
$e^\pm$. The only difference is that we adopt the intensities of incident H and
He given by Ref. \cite{mauro14}. Besides, a rescaling parameter $c_{\rm e}$ is
introduced for the secondary component in the following fitting procedure,
considering the uncertainties in the calculation.

\section{Explanation of the DAMPE Data and Fitting to the Model Parameters}
\label{sec:results}
In this section, we show our explanation of the DAMPE data and fitting results of
the model parameters.
Here we repeat again all the components needed to account for
the total $e^\pm$ spectrum: the population
A and population B electrons from SNRs, a single pulsar as $e^\pm$ source,
and secondary $e^\pm$.
The common free parameters are
$Q_{0, {\rm A}}$, $\gamma_{\rm A}$, $t_{\rm end}$, and $c_{\rm e}$ which
are determined by the fitting to the DAMPE $e^-+e^+$ and AMS-02 $e^+/(e^-+e^+)$ data.
We seek the
best-fit model parameters by minimizing chi-squared statics\footnote{Steven G.
Johnson, The NLopt nonlinear-optimization package,
http://ab-initio.mit.edu/nlopt} between model predictions and the experimental data. The best-fit
parameters of these models are compiled in Table \ref{tab:psr}.

\begin{table}[t]
 \center
\begin{tabular}{ccccc}
 \hline
 Single pulsar & & \multicolumn{2}{c}{J0940} & Geminga \\
 \hline
 With Pop B? & & No & Yes & Yes \\
 \hline
 Parameters & Bounds & Best-fit & Best-fit & Best-fit \\
 \hline
 $Q_{0, {\rm A}}$ [$10^{50}$ GeV$^{-1}$] & [0.1, 10.0] & 2.60 & 2.61 & 2.66 \\
 $\gamma_{\rm A}$ & [2.0, 3.0] & 2.497 & 2.498 & 2.501 \\
 $t_{\rm end}$ [$10^5$ yr] & [0.4, 5.0] & 0.94 & 0.89 & 1.44 \\
 $\dot{Q}_{0, {\rm B}}$ [$10^{33}$ GeV$^{-1}$ s$^{-1}$] & [0.0, 10.0] & - &
1.72 & 0.19 \\
 $\eta_{\rm pwn}$ & [0.0, 2.0] & 1.29 & 1.21 & 1.26 \\
 $\gamma_{\rm pwn}$ & [1.0, 2.7] & 2.54 & 2.52 & 1.96 \\
 $E_{\rm c, pwn}$ [TeV] & [0.8, 10.0] & 3.24 & 2.37 & - \\
 $c_{\rm e}$ & [0.5, 2.0] & 1.41& 1.42 & 0.5 \\
 \hline
 $\chi^2$/d.o.f & & 0.72 & 0.71 & 0.81 \\
 \hline
\end{tabular}
\caption{Best-fit parameters for three cases. The reduced chi-square of
each model is also shown, and 'd.o.f' is the abbreviation of degree of freedom.
As described in Sec. \ref{subsubsec:psr}, the cut-off energy of Geminga is
fixed at 50 TeV, so its best-fit value of $E_{\rm c,pwn}$ is blank.}
\label{tab:psr}
\end{table}

\begin{figure}[t]
\centering
\includegraphics[width=0.35\textwidth, angle=270]{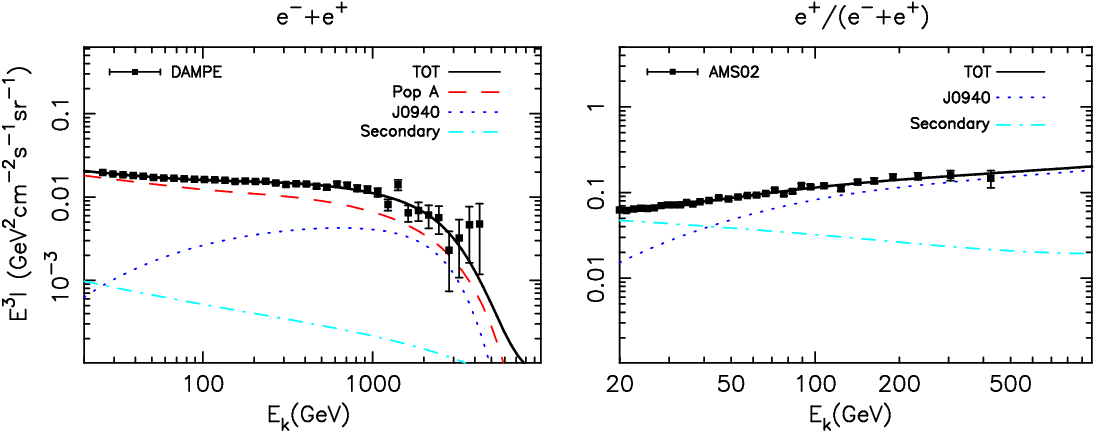}
\includegraphics[width=0.35\textwidth, angle=270]{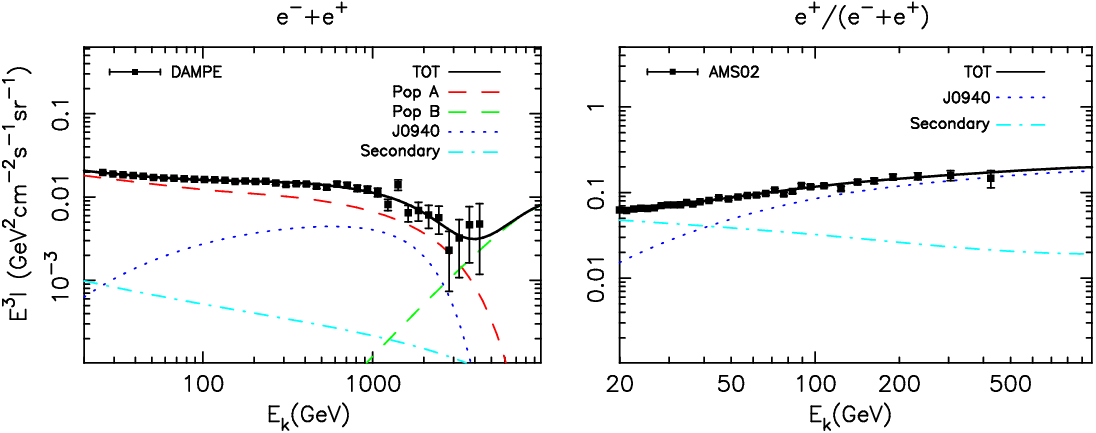}
\caption{Results of the fitting to the DAMPE $e^-+e^+$ and AMS-02 $e^+/(e^-+e^+)$
data, with PSR J0940-5428 as the single positron source in the high
energy region. Left panels: the electron plus positron spectrum. Right panels:
the positron fraction. The results in the bottom panels include the
population B of SNRs, while the results in the top panels do not. In the legends,
'TOT' stands for the total $e^-+e^+$ flux (or $e^+/(e^-+e^+)$) of all the
sources, the population A of SNRs are abbreviated to 'Pop A', the population B are
abbreviated to 'Pop B', while 'J0940' stands for PSR J0940-5428. }
\label{fig:J0940}
\end{figure}

\begin{figure}[t]
\centering
\includegraphics[width=0.35\textwidth, angle=270]{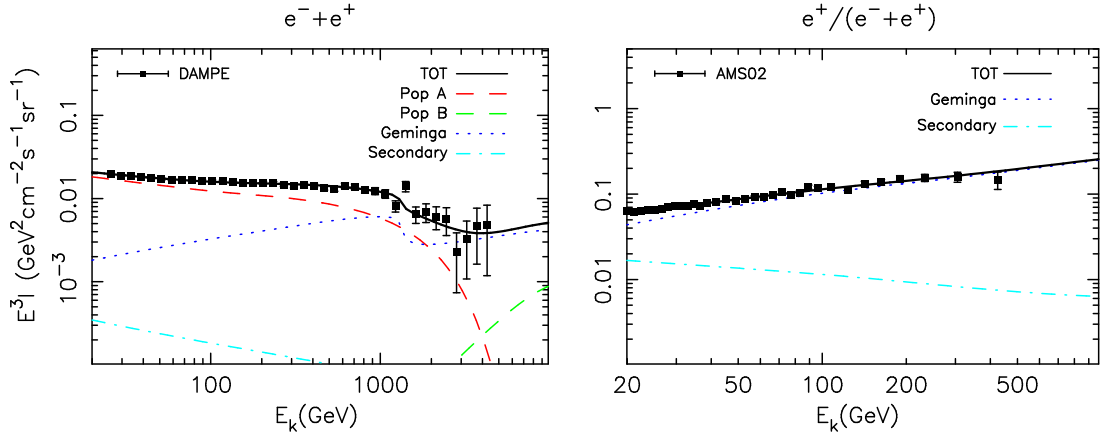}
\caption{Same as Fig. \ref{fig:J0940}, but for the case of Geminga as
the high energy positron source.}
\label{fig:geminga}
\end{figure}

Figs. \ref{fig:J0940} and \ref{fig:geminga} show the fitting results compared
with the DAMPE and AMS-02 data. As expected, the TeV break of the
DAMPE $e^-+e^+$
spectrum can be well reproduced in all cases, owing to the cut-off
energy of the population A of SNRs. The best-fit $t_{\rm end}$ varies
from $\sim9\times10^4$ yr to $\sim1.4\times10^5$ yr,
which is quite reasonable for the final release time of particles in SNRs. The
corresponding cut-off energies of injection spectra of the population A vary from 3.5 TeV to 2.5 TeV.

In Figs. \ref{fig:J0940} and \ref{fig:geminga} we take the pulsar
PSR J0940-5428 and Geminga as the single positron source respectively.
As shown in the figures, the $e^-+e^+$ spectra of PSR J0940-5428 and Geminga are significantly different
in the TeV region, which can be attributed to the differences between their
$t$ and $E_c$. The spectral break of Geminga around 1 TeV is due to its old age:
under the assumption of Eq. \ref{eq:Q_p}, most spin-down energy of the pulsar is released in the form of $e^\pm$ in the early age of Geminga, and the high energy part of the $e^\pm$ has been effectively cooled by now. On the other hand, the very high Ec ensures the spectrum of Geminga to extend to tens of TeV, which is contributed by the $e^\pm$ injected at the later age of Geminga.
This indicates that no significant contribution from the population B is needed to prevent a fast spectral cut-off at the high energy end of DAMPE, as shown in Fig. \ref{fig:geminga}.

In comparison, since PSR J0940-5428 is a much younger source, the plenty of high
energy $e^\pm$ released in its early age are still energetic now. The fitting result requires a
large cut-off energy of $\sim3$ TeV to suppress the spectrum at high energies
and produce the break feature at $\sim$ TeV. In this case, the values of
$E^3I(E)$ for the population A and the single pulsar descend in several TeV. Therefore
the total $e^-+e^+$ spectrum quickly rolls off at higher energies, as shown in the
left panel of Fig. \ref{fig:J0940}. If we expect an additional spectral feature in the TeV region, the
population B of SNRs is needed. The best-fit injection rate of the
population B is $\dot{Q}_{0, {\rm B}}=1.72\times10^{33}$
GeV$^{-1}$ s$^{-1}$, corresponding to a total injection energy of $\sim10^{46}$
erg for an SNR with an age of $10^4$ yr (like Vela). Since the typical kinetic energy
of SN ejecta is $10^{51}$ erg, only a fraction of $\sim10^{-5}$ of the total
energy is required to deliver to the population B.

We notice that the conversion efficiencies of the total spin-down energy to
positrons are larger than 1 for the two single pulsar cases. However, this is not a crucial problem. On
the one hand, the spin-down time scale $\tau_0$ is quite uncertain, which is
important for the estimation of the spin-down energy. On the other hand, the
single pulsar model is definitely a simplified picture. In the realistic case,
$e^\pm$ should be contributed by more pulsars, which may alleviate the demand
for a large $\eta_{\rm pwn}$.

The fitting requires a reasonable $\gamma_{\rm pwn}$ of 1.96 for Geminga, while
the best-fit $\gamma_{\rm pwn}$ for PSR J0940-5428 is $\sim2.5$. In fact, among all the
identified PWNe with measurements of radio spectral indices, only DA 495 has an
electron spectral index larger than 2.5 (to be specific,
2.74)\cite{2008ApJ...687..516K,2017SSRv..207..175R}. The requirement of such a soft injection spectrum for PSR J0940-5428 in the fitting
is attributed to its younger age and farther distance,
compared with those of Geminga. In general, a source with a large $r$
and small $t$ yields
a hard $e^\pm$ spectrum at the Earth, which cannot well fit
the $e^+/(e^-+e^+)$ fraction of AMS-02. If the source accommodates the
$e^+/(e^-+e^+)$ data at low energies, the hard positron spectrum would
cause a deviation from the AMS-02 data above $\sim 300$ GeV. Hence a large $\gamma_{\rm
pwn}$ and soft injection spectrum are required for PSR J0940-5428 as the single source to explain the
positron observation.

Finally, we briefly discuss the tentative spectral feature around 1.4 TeV.
If this feature is a sharp peak, a mono-energetic electron injection at 1.4 TeV from a nearby source is required, such as dark matter annihilation in a subhalo.
However, if there exists some unclear uncertainties, the spectral feature would not be a strictly peak. For instance, if the count number at $\sim1.2$ TeV is underestimated, the actual spectral feature should be edge-like around 1.4 TeV.
We have described above that the spectrum of Geminga breaks around $\sim1.2$ TeV due to cooling effect, which can create an edge-like feature at the TeV region, as shown in Fig.~\ref{fig:geminga}. The precise property of this tentative spectral feature should be tested with larger statistics of further results of DAMPE.

\section{A Powerful Nearby SNR as the Dominant High Energy Electron Source}
\label{sec:mr}
In the previous sections, the population A is assumed to be contributed by smoothly
distributed sources. In fact, electrons with $E>1$ TeV may be mainly contributed by sources
with $r<1.5$ kpc and $t_0<3\times10^5$ yr due to the radiative
cooling of the travelling electrons. This condition can be derived by the
criteria given in Appendix B of Ref. \cite{koba04}. For $t_{\rm end}\simeq10^5$
yr, among observed SNRs, only the Monogem
Ring (MR) and Loop I can satisfy this condition. Other observed SNRs within 1.5 kpc are too young to contribute to the
population A. In this section, we show the scenario in which a powerful
nearby SNR dominantly accounts for the TeV spectrum of the population A, and choose MR as an example. We take the distance of MR to be $r_{\rm mr}=300$ pc
\cite{plucinsky09}, and set its age to be $t_{\rm mr}=1.1\times10^5$ yr,
which is same as its associated pulsar PSR B0656+14 \cite{atnf}.

\begin{figure}[t]
\centering
\includegraphics[width=0.35\textwidth, angle=270]{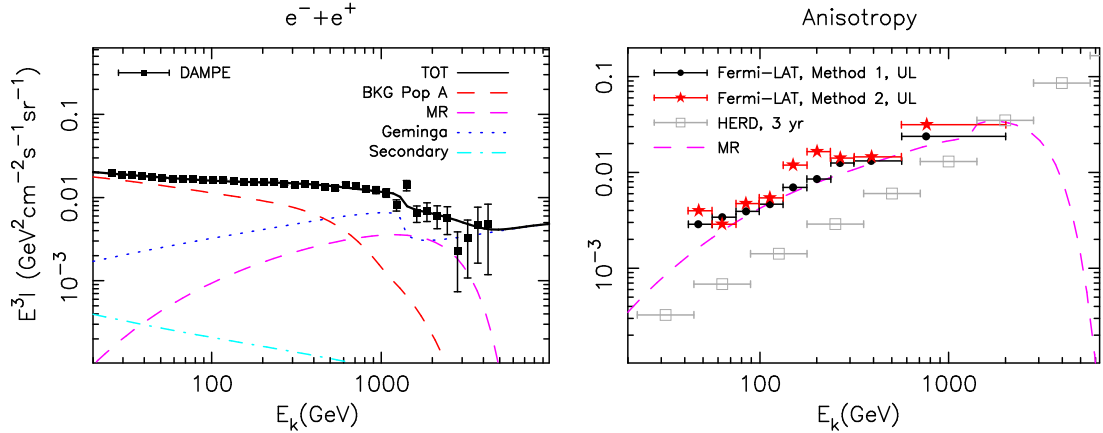}
\caption{Left panel: The $e^-+e^+$ spectrum of the model where the
population A are contributed by a nearby source MR and a smooth distributed
SNR background (labeled with 'BKG Pop A'). The best-fit parameters are:
$\gamma_{\rm mr}=2.1$, $W_{\rm mr}=8.8\times10^{48}$ erg, $Q_{0, {\rm
A}}=1.92\times10^{50}$
GeV$^{-1}$, $\gamma_{\rm A}=2.36$, $t_{\rm end}=9\times10^4$ yr, $\gamma_{\rm
pwn}=1.93$, $\eta_{\rm pwn}=1.23$, and $c_{\rm e}=0.57$. Right panel: the
corresponding anisotropy spectrum, assuming that the MR contribution is dominant. The anisotropy is calculated by $3r_{\rm mr}/[2c(t_{\rm mr}-t_{\rm
end})]$ multiplying the relative flux of MR. The two groups upper limits (UL)
of Fermi-LAT are given by the two different analysis methods in Ref.~\cite{2017PhRvL.118i1103A}. An expectation of the sensitivity of HERD is also
shown.}
\label{fig:mr}
\end{figure}

To calculate the electron spectrum, we consider the population A contributed by
MR and a smooth distributed SNR background. We adopt the framework in Ref.
\cite{2017ApJ...836..172F}, but with the electron injection
introduced in the present work. The injection age of MR is $t_{\rm mr}-t_{\rm
end}$; the cut-off energy of its injection spectrum $E_c(t_{\rm end})$ is
estimated by Eq. (\ref{eq:p_c}) and (\ref{eq:u1_3}). The spectral index of MR
$\gamma_{\rm mr}$ and total energy of injected electrons $W_{\rm mr}$ are
both assumed to be free. Then we perform the fitting to the DAMPE $e^-+e^+$
spectrum and the positron fraction of AMS-02 again, adopting Geminga as the single high
energy positron source. The population B of SNRs is not included here. The fitting
result is presented in the left panel of Fig. \ref{fig:mr}; the best-fit
parameters can be found in the caption. The positron fraction is similar to
that in the right panel of Fig. \ref{fig:geminga}, and thus is not shown here.

Similar to the result in Fig. \ref{fig:geminga}, this scenario can also well fit the DAMPE
data. The best-fit spectral index $\gamma_{\rm mr}=2.1$ and total
injection energy $W_{\rm mr}=8.8\times10^{48}$ erg are reasonable for an
SNR. The spectral shape at $\sim1$ TeV in the model is mainly determined by the properties of MR
and Geminga. Thus the TeV break of DAMPE can be coincidentally generated by one
or few powerful nearby sources, as indicated by the previous works
\cite{1995PhRvD..52.3265A,mauro14,2017ApJ...836..172F}.

Meanwhile, nearby sources with young injection ages can produce significant anisotropy of $e^-+e^+$, as the dipole anisotropy of a burst-like point source can be roughly estimated by $3r/(2ct_0)$ \cite{1971ApL.....9..169S}. The detection of the anisotropy may help to ascertain the contribution of high energy $e^-+e^+$ \cite{2013ApJ...772...18L,2017JCAP...01..006M,2017arXiv170603745F}. Since the injection age of MR is only $2\times10^4$ yr, we show the case that MR dominates the $e^-+e^+$ anisotropy of the model in the right panel Fig. \ref{fig:mr}. We compare the predicted anisotropy with the upper limits given by the seven years detection of Fermi-LAT \cite{2017PhRvL.118i1103A}. The anisotropy produced by MR almost hits the upper limits of Fermi-LAT. Future instruments with larger acceptance and stronger
electron/proton discrimination, like HERD \cite{2014SPIE.9144E..0XZ}, will
provide further judgement to a scenario discussed in this section. An expectation of the
sensitivity of HERD with three years measurement is shown in the right panel of
Fig. \ref{fig:mr}; one may refer to Ref. \cite{2017arXiv170603745F} for the
estimation of the sensitivity.

\section{Conclusion}
\label{sec:conclusion}

In this work, we apply an electron escaping scenario of SNRs to explain the DAMPE $e^-+e^+$ spectrum. In this scenario, most accelerated electrons are released at the final stage of the evolution of SNRs, while a minor part of electrons may escape during their acceleration process. The similar scenario has already been utilized in the researches of CR protons. We also explain the $e^+/(e^-+e^+)$ fraction measured by AMS-02 at the same time, using a single powerful pulsar as the primary high energy positron source. The main conclusions are presented as follows.

\begin{enumerate}
\item The spectral break at $\sim0.9$ TeV can be naturally explained by the final released electrons of smoothly distributed SNRs, which escape at an age of $\sim 10^5$ yr.
\item The break can also be mainly induced by a single nearby old SNR, this case will be hopefully tested by future measurements of the anisotropy of $e^-+e^+$ flux with future instruments, such as HERD.
\item A minor part of electrons escaping during the evolution of nearby SNRs are able to contribute to the TeV region significantly, owing to the hard spectral index of the injection spectrum.
\item The single pulsar, such as Geminga, can be a dominant source of high energy CR positrons. The sharp drop at $\sim 1$ TeV of the $e^\pm$ spectrum from Geminga can induce an edge-like feature at the $e^-+e^+$ spectrum, which may account for the tentative feature of the DAMPE data at $\sim$1.4 TeV.
\end{enumerate}

\acknowledgments{We thank Qiang Yuan for very helpful discussions.
This work is supported by the National Key Program for Research and Development (No.~2016YFA0400200) and by the National Natural Science Foundation of China under Grants No.~11475189,~11475191.}

\appendix

\section{Electrons Injected from a Free-escape Boundary}
\label{app:escape}
To describe the injection of electrons during
the acceleration, we impose the free boundary condition of $f(-x_b)=0$ at a position $-x_b$
in the upstream of the SNR shock \cite{2010APh....33..307C}.
At low energies, the distribution of electrons in the
upstream can be simply described by the diffusion-convection
equation. The solution is expressed as
\begin{equation}
 f(x,p)=f_0(p)\,\frac{{\rm exp}(u_1x/D_1)-{\rm exp}(-u_1x_b/D_1)}{1-{\rm
exp}(-u_1x_b/D_1)}\,.
\label{eq:f_u}
\end{equation}
Then we can get the escape flux $F_{\rm esc}=D_1\partial f/\partial x|_{x_b}$
in the whole energy range as
\begin{equation}
 F_{\rm esc}=\frac{u_1f_0(p)}{{\rm exp}(u_1x_b/D_1)-1}\,,
\end{equation}
where $f_0(p)\propto p^{-4}{\rm exp}(-p^2/p_c^2)$. The free-escape boundary is
placed at $x_b=cR_{sh}$, where $R_{sh}$ is the radius of the SNR and $c$ is a
constant \cite{2010APh....33..160C}. The radius $R_{sh}$ is derived
by Eq. (\ref{eq:u1_2}) and Eq. (\ref{eq:u1_3}), and the injection spectrum
should be proportional to $F_{\rm esc}R_{sh}^2$. Here we take $c$ to be 0.01.
For a larger
$c$, particles escaping from the free-escape boundary are required to be
much more energetic than the cut-off energy due to radiative cooling. This
means that few particles could escape for a larger $c$.

\begin{figure}[t]
\centering
\includegraphics[width=0.35\textwidth, angle=270]{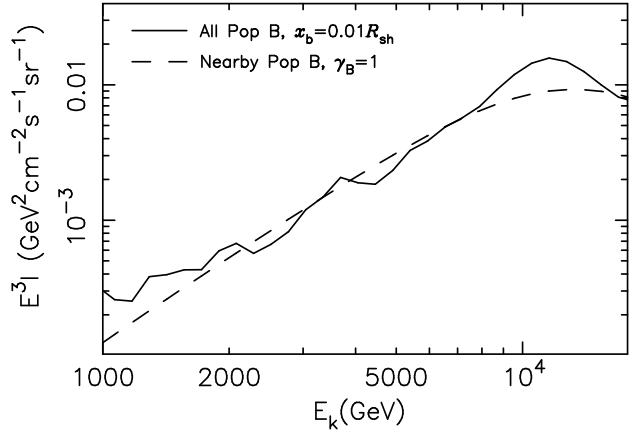}
\caption{The electron spectrum of the population B contributed by a complete group
of SNRs with a free-escape boundary (solid line), compared with the spectrum of the
population B shown in the top left panel of Fig. \ref{fig:J0940} (dashed
line).}
\label{fig:freeb}
\end{figure}

In the main text, the sources of the population B consist of only young
and nearby SNRs. Here we would like to build a complete group of the sources
of the population B. Taking the birth rate of SNe of
4 century$^{-1}$ galaxy$^{-1}$ and the radial distribution of SNRs given
in Ref. \cite{l04}, we generate $4\times10^5$ SNRs randomly with the age from
0 to $10^7$ yr (SNRs older than $10^7$ yr can hardly contribute to the energy
range in which we are interested). We calculate the injection
spectra for all the sources, and use Eq. (\ref{eq:I_c})
to obtain the observed spectrum at the Earth. The final spectrum generated with this method is drawn in Fig.
\ref{fig:freeb}, compared with the spectrum of the population B shown in the top left
panel of Fig. \ref{fig:J0940}. The normalization of the generated injection spectrum has been
adjusted to produce a comparable flux with that shown in Fig. \ref{fig:J0940}.

Although the contributions of distant or/and old SNRs to the population B are included, the model described
in this section generates a similar spectral shape with that predicted by the model in Sec.~\ref{subsubsec:pop_B}, where only young and nearby
sources are considered and the integrated upstream spectrum is adopted as the injection spectrum.
This is because that the injection spectra from SNRs with a free-escape boundary described
in this section are much harder.

\section{The Injection Spectrum of Pulsars}
\label{app:pulsar}
Pulsars convert their spin-down energy partially to relativistic winds of
$e^\pm$ pairs. The young or middle aged pulsar may be surrounded by an
observable PWN (see Ref. \cite{pwn06} and references
therein). Particles injected into the PWN are constrained in a period of time
until the crush of the PWN, and then escape into the ISM
\cite{2004A&A...420..937V}. Thus the injection spectrum of $e^\pm$ should be
the spectrum from the PWN, rather than that from the pulsar itself \cite{maly09}.
The particle acceleration and escape processes in PWNe are much more complicated than those
in SNRs. Here we simply assume that the injection process has the same time
dependency with the spin-down luminosity of pulsars (i.e.,
$\propto(1+t/\tau_0)^{-2}$ \cite{1973ApJ...186..249P}), and the $e^\pm$
injection spectrum of PWNe takes the form of
\begin{equation}
 Q(E, t_0)=C\,[1+(t-t_0)/\tau_0]^{-2}\,{(E/{\rm 1\,GeV})}^{-\gamma}{\rm
exp}(-E/E_c)\,,
 \label{eq:Q_p}
\end{equation}
where $C$ is the normalization, $t$ is the age of a pulsar, and $t_0$ denotes
the injection time. Meanwhile, $Q$ can be related to the total spin-down energy
released by a pulsar by
\begin{equation}
\int_{E_{\rm min}}^{\infty}dE\int_{r/c}^{t}dt_0\,Q(E,t_0)E=\eta W_{\rm p}\,,
\label{eq:pwn}
\end{equation}
where $W_{\rm p}$ is the spin-down energy, $\eta$ represents the conversion
efficiency to injected positrons, $r$ is the distance of
the pulsar, and $E_{\rm min}$ is set to be 0.1 GeV. We integrate the spin-down
luminosity $\dot{E}$ with time to get the spin-down energy
\begin{equation}
 W_{\rm p}=\dot{E}\,t\,(1+\frac{t}{\tau_0})\,,
 \label{eq:W_p}
\end{equation}
where $\tau_0$ is the initial spin-down timescale of the pulsar, which has a
typical value of $10$ kyr \cite{2017arXiv170208436H}. $\dot{E}$
and $t$ of known pulsars are given by the ATNF catalog \citep{2005AJ....129.1993M}, so for each pulsar we
can perform an estimation to $W_{\rm p}$. Then the injection spectrum of a pulsar
is determined by three parameters: $\eta$, $\gamma$, and $E_{\rm c}$.

So far, the ATNF catalog has collected 2573 pulsars. We gave a rank of all
these pulsars by their potential of contribution to the arrival $e^\pm$ spectrum
in Ref. \cite{2017arXiv170603745F}. One may refer to Ref.
\cite{2017arXiv170603745F} for the rule of this rank. Vela pulsar and
PSR J0940-5428 are much more powerful in $e^\pm$ spectrum than other pulsars.
So if we intend to reproduce the positron excess by a single pulsar, these two
sources may require the smallest $\eta$. However, as we have described above,
$e^\pm$ may not be injected from a PWN
until the crush of the PWN. It may take thousands of years for the reverse
shock of an SNR to propagate backward and collide with its PWN
\cite{blondin01,hinton11}. Since Vela pulsar has an age of $1.1\times10^4$
years, an injection delay of thousands of years may push it to contribute in
higher energy region of $e^\pm$ spectrum. While for PSR J0940-5428, which is
$4.2\times10^4$ years old, the injection delay may have less influence.
Therefore, we choose PSR J0940-5428 as the single positron source in this work.

\bibliography{dampe}

\begin{thebibliography}{62}%
\makeatletter
\providecommand \@ifxundefined [1]{%
 \@ifx{#1\undefined}
}%
\providecommand \@ifnum [1]{%
 \ifnum #1\expandafter \@firstoftwo
 \else \expandafter \@secondoftwo
 \fi
}%
\providecommand \@ifx [1]{%
 \ifx #1\expandafter \@firstoftwo
 \else \expandafter \@secondoftwo
 \fi
}%
\providecommand \natexlab [1]{#1}%
\providecommand \enquote  [1]{``#1''}%
\providecommand \bibnamefont  [1]{#1}%
\providecommand \bibfnamefont [1]{#1}%
\providecommand \citenamefont [1]{#1}%
\providecommand \href@noop [0]{\@secondoftwo}%
\providecommand \href [0]{\begingroup \@sanitize@url \@href}%
\providecommand \@href[1]{\@@startlink{#1}\@@href}%
\providecommand \@@href[1]{\endgroup#1\@@endlink}%
\providecommand \@sanitize@url [0]{\catcode `\\12\catcode `\$12\catcode
  `\&12\catcode `\#12\catcode `\^12\catcode `\_12\catcode `\%12\relax}%
\providecommand \@@startlink[1]{}%
\providecommand \@@endlink[0]{}%
\providecommand \url  [0]{\begingroup\@sanitize@url \@url }%
\providecommand \@url [1]{\endgroup\@href {#1}{\urlprefix }}%
\providecommand \urlprefix  [0]{URL }%
\providecommand \Eprint [0]{\href }%
\providecommand \doibase [0]{http://dx.doi.org/}%
\providecommand \selectlanguage [0]{\@gobble}%
\providecommand \bibinfo  [0]{\@secondoftwo}%
\providecommand \bibfield  [0]{\@secondoftwo}%
\providecommand \translation [1]{[#1]}%
\providecommand \BibitemOpen [0]{}%
\providecommand \bibitemStop [0]{}%
\providecommand \bibitemNoStop [0]{.\EOS\space}%
\providecommand \EOS [0]{\spacefactor3000\relax}%
\providecommand \BibitemShut  [1]{\csname bibitem#1\endcsname}%
\let\auto@bib@innerbib\@empty
\bibitem [{\citenamefont {{DAMPE Collaboration}}(2017)}]{nature}%
  \BibitemOpen
  \bibfield  {author} {\bibinfo {author} {\bibnamefont {{DAMPE
  Collaboration}}},\ }\href@noop {} {\bibfield  {journal} {\bibinfo  {journal}
  {Nature in press}\ ,\ \bibinfo {pages} {doi:10.1038/nature24475}} (\bibinfo
  {year} {2017})}\BibitemShut {NoStop}%
\bibitem [{\citenamefont {{Chang}}(2014)}]{dampe}%
  \BibitemOpen
  \bibfield  {author} {\bibinfo {author} {\bibfnamefont {J.}~\bibnamefont
  {{Chang}}},\ }\href@noop {} {\bibfield  {journal} {\bibinfo  {journal} {Chin.
  J. Space Sci.}\ }\textbf {\bibinfo {volume} {34}},\ \bibinfo {pages} {550}
  (\bibinfo {year} {2014})}\BibitemShut {NoStop}%
\bibitem [{\citenamefont {{Aharonian}}\ \emph {et~al.}(2009)\citenamefont
  {{Aharonian}} \emph {et~al.}}]{hess09}%
  \BibitemOpen
  \bibfield  {author} {\bibinfo {author} {\bibfnamefont {F.}~\bibnamefont
  {{Aharonian}}} \emph {et~al.},\ }\href {\doibase 10.1051/0004-6361/200913323}
  {\bibfield  {journal} {\bibinfo  {journal} {\aap}\ }\textbf {\bibinfo
  {volume} {508}},\ \bibinfo {pages} {561} (\bibinfo {year} {2009})},\ \Eprint
  {http://arxiv.org/abs/0905.0105} {arXiv:0905.0105 [astro-ph.HE]} \BibitemShut
  {NoStop}%
\bibitem [{\citenamefont {{Staszak}}\ and\ \citenamefont {{for the VERITAS
  Collaboration}}(2015)}]{veritas}%
  \BibitemOpen
  \bibfield  {author} {\bibinfo {author} {\bibfnamefont {D.}~\bibnamefont
  {{Staszak}}}\ and\ \bibinfo {author} {\bibnamefont {{for the VERITAS
  Collaboration}}},\ }\href@noop {} {\bibfield  {journal} {\bibinfo  {journal}
  {ArXiv e-prints}\ } (\bibinfo {year} {2015})},\ \Eprint
  {http://arxiv.org/abs/1508.06597} {arXiv:1508.06597 [astro-ph.HE]}
  \BibitemShut {NoStop}%
\bibitem [{\citenamefont {Amenomori}\ \emph {et~al.}(2011)\citenamefont
  {Amenomori} \emph {et~al.}}]{Amenomori:2011zza}%
  \BibitemOpen
  \bibfield  {author} {\bibinfo {author} {\bibfnamefont {M.}~\bibnamefont
  {Amenomori}} \emph {et~al.} (\bibinfo {collaboration} {Tibet ASgamma}),\
  }\href {\doibase 10.1016/j.asr.2010.08.029} {\bibfield  {journal} {\bibinfo
  {journal} {Adv. Space Res.}\ }\textbf {\bibinfo {volume} {47}},\ \bibinfo
  {pages} {629} (\bibinfo {year} {2011})}\BibitemShut {NoStop}%
\bibitem [{\citenamefont {Bartoli}\ \emph {et~al.}(2015)\citenamefont {Bartoli}
  \emph {et~al.}}]{Bartoli:2015vca}%
  \BibitemOpen
  \bibfield  {author} {\bibinfo {author} {\bibfnamefont {B.}~\bibnamefont
  {Bartoli}} \emph {et~al.} (\bibinfo {collaboration} {LHAASO, ARGO-YBJ}),\
  }\href {\doibase 10.1103/PhysRevD.92.092005} {\bibfield  {journal} {\bibinfo
  {journal} {Phys. Rev.}\ }\textbf {\bibinfo {volume} {D92}},\ \bibinfo {pages}
  {092005} (\bibinfo {year} {2015})},\ \Eprint
  {http://arxiv.org/abs/1502.03164} {arXiv:1502.03164 [astro-ph.HE]}
  \BibitemShut {NoStop}%
\bibitem [{\citenamefont {{H{\"o}randel}}(2004)}]{2004APh....21..241H}%
  \BibitemOpen
  \bibfield  {author} {\bibinfo {author} {\bibfnamefont {J.~R.}\ \bibnamefont
  {{H{\"o}randel}}},\ }\href {\doibase 10.1016/j.astropartphys.2004.01.004}
  {\bibfield  {journal} {\bibinfo  {journal} {Astroparticle Physics}\ }\textbf
  {\bibinfo {volume} {21}},\ \bibinfo {pages} {241} (\bibinfo {year} {2004})},\
  \Eprint {http://arxiv.org/abs/astro-ph/0402356} {astro-ph/0402356}
  \BibitemShut {NoStop}%
\bibitem [{\citenamefont {{Baade}}\ and\ \citenamefont
  {{Zwicky}}(1934)}]{1934PNAS...20..259B}%
  \BibitemOpen
  \bibfield  {author} {\bibinfo {author} {\bibfnamefont {W.}~\bibnamefont
  {{Baade}}}\ and\ \bibinfo {author} {\bibfnamefont {F.}~\bibnamefont
  {{Zwicky}}},\ }\href {\doibase 10.1073/pnas.20.5.259} {\bibfield  {journal}
  {\bibinfo  {journal} {Proceedings of the National Academy of Science}\
  }\textbf {\bibinfo {volume} {20}},\ \bibinfo {pages} {259} (\bibinfo {year}
  {1934})}\BibitemShut {NoStop}%
\bibitem [{\citenamefont {{Ginzburg}}\ and\ \citenamefont
  {{Syrovatsky}}(1961)}]{1961PThPS..20....1G}%
  \BibitemOpen
  \bibfield  {author} {\bibinfo {author} {\bibfnamefont {V.~L.}\ \bibnamefont
  {{Ginzburg}}}\ and\ \bibinfo {author} {\bibfnamefont {S.~I.}\ \bibnamefont
  {{Syrovatsky}}},\ }\href {\doibase 10.1143/PTPS.20.1} {\bibfield  {journal}
  {\bibinfo  {journal} {Progress of Theoretical Physics Supplement}\ }\textbf
  {\bibinfo {volume} {20}},\ \bibinfo {pages} {1} (\bibinfo {year}
  {1961})}\BibitemShut {NoStop}%
\bibitem [{\citenamefont {{Blasi}}(2013)}]{2013A&ARv..21...70B}%
  \BibitemOpen
  \bibfield  {author} {\bibinfo {author} {\bibfnamefont {P.}~\bibnamefont
  {{Blasi}}},\ }\href {\doibase 10.1007/s00159-013-0070-7} {\bibfield
  {journal} {\bibinfo  {journal} {\aapr}\ }\textbf {\bibinfo {volume} {21}},\
  \bibinfo {eid} {70} (\bibinfo {year} {2013})},\ \Eprint
  {http://arxiv.org/abs/1311.7346} {arXiv:1311.7346 [astro-ph.HE]} \BibitemShut
  {NoStop}%
\bibitem [{\citenamefont {{Delahaye}}\ \emph {et~al.}(2010)\citenamefont
  {{Delahaye}}, \citenamefont {{Lavalle}}, \citenamefont {{Lineros}},
  \citenamefont {{Donato}},\ and\ \citenamefont {{Fornengo}}}]{dela10}%
  \BibitemOpen
  \bibfield  {author} {\bibinfo {author} {\bibfnamefont {T.}~\bibnamefont
  {{Delahaye}}}, \bibinfo {author} {\bibfnamefont {J.}~\bibnamefont
  {{Lavalle}}}, \bibinfo {author} {\bibfnamefont {R.}~\bibnamefont
  {{Lineros}}}, \bibinfo {author} {\bibfnamefont {F.}~\bibnamefont {{Donato}}},
  \ and\ \bibinfo {author} {\bibfnamefont {N.}~\bibnamefont {{Fornengo}}},\
  }\href {\doibase 10.1051/0004-6361/201014225} {\bibfield  {journal} {\bibinfo
   {journal} {\aap}\ }\textbf {\bibinfo {volume} {524}},\ \bibinfo {eid} {A51}
  (\bibinfo {year} {2010})},\ \Eprint {http://arxiv.org/abs/1002.1910}
  {arXiv:1002.1910 [astro-ph.HE]} \BibitemShut {NoStop}%
\bibitem [{\citenamefont {{Di Mauro}}\ \emph {et~al.}(2014)\citenamefont {{Di
  Mauro}}, \citenamefont {{Donato}}, \citenamefont {{Fornengo}}, \citenamefont
  {{Lineros}},\ and\ \citenamefont {{Vittino}}}]{mauro14}%
  \BibitemOpen
  \bibfield  {author} {\bibinfo {author} {\bibfnamefont {M.}~\bibnamefont {{Di
  Mauro}}}, \bibinfo {author} {\bibfnamefont {F.}~\bibnamefont {{Donato}}},
  \bibinfo {author} {\bibfnamefont {N.}~\bibnamefont {{Fornengo}}}, \bibinfo
  {author} {\bibfnamefont {R.}~\bibnamefont {{Lineros}}}, \ and\ \bibinfo
  {author} {\bibfnamefont {A.}~\bibnamefont {{Vittino}}},\ }\href {\doibase
  10.1088/1475-7516/2014/04/006} {\bibfield  {journal} {\bibinfo  {journal}
  {\jcap}\ }\textbf {\bibinfo {volume} {4}},\ \bibinfo {eid} {006} (\bibinfo
  {year} {2014})},\ \Eprint {http://arxiv.org/abs/1402.0321} {arXiv:1402.0321
  [astro-ph.HE]} \BibitemShut {NoStop}%
\bibitem [{\citenamefont {{Caprioli}}\ \emph
  {et~al.}(2010{\natexlab{a}})\citenamefont {{Caprioli}}, \citenamefont
  {{Amato}},\ and\ \citenamefont {{Blasi}}}]{2010APh....33..160C}%
  \BibitemOpen
  \bibfield  {author} {\bibinfo {author} {\bibfnamefont {D.}~\bibnamefont
  {{Caprioli}}}, \bibinfo {author} {\bibfnamefont {E.}~\bibnamefont {{Amato}}},
  \ and\ \bibinfo {author} {\bibfnamefont {P.}~\bibnamefont {{Blasi}}},\ }\href
  {\doibase 10.1016/j.astropartphys.2010.01.002} {\bibfield  {journal}
  {\bibinfo  {journal} {Astroparticle Physics}\ }\textbf {\bibinfo {volume}
  {33}},\ \bibinfo {pages} {160} (\bibinfo {year} {2010}{\natexlab{a}})},\
  \Eprint {http://arxiv.org/abs/0912.2964} {arXiv:0912.2964 [astro-ph.HE]}
  \BibitemShut {NoStop}%
\bibitem [{\citenamefont {{Shen}}(1970)}]{shen70}%
  \BibitemOpen
  \bibfield  {author} {\bibinfo {author} {\bibfnamefont {C.~S.}\ \bibnamefont
  {{Shen}}},\ }\href {\doibase 10.1086/180650} {\bibfield  {journal} {\bibinfo
  {journal} {\apjl}\ }\textbf {\bibinfo {volume} {162}},\ \bibinfo {pages}
  {L181} (\bibinfo {year} {1970})}\BibitemShut {NoStop}%
\bibitem [{\citenamefont {{Kobayashi}}\ \emph {et~al.}(2004)\citenamefont
  {{Kobayashi}}, \citenamefont {{Komori}}, \citenamefont {{Yoshida}},\ and\
  \citenamefont {{Nishimura}}}]{koba04}%
  \BibitemOpen
  \bibfield  {author} {\bibinfo {author} {\bibfnamefont {T.}~\bibnamefont
  {{Kobayashi}}}, \bibinfo {author} {\bibfnamefont {Y.}~\bibnamefont
  {{Komori}}}, \bibinfo {author} {\bibfnamefont {K.}~\bibnamefont {{Yoshida}}},
  \ and\ \bibinfo {author} {\bibfnamefont {J.}~\bibnamefont {{Nishimura}}},\
  }\href {\doibase 10.1086/380431} {\bibfield  {journal} {\bibinfo  {journal}
  {\apj}\ }\textbf {\bibinfo {volume} {601}},\ \bibinfo {pages} {340} (\bibinfo
  {year} {2004})},\ \Eprint {http://arxiv.org/abs/astro-ph/0308470}
  {astro-ph/0308470} \BibitemShut {NoStop}%
\bibitem [{\citenamefont {{Adriani}}\ \emph {et~al.}(2013)\citenamefont
  {{Adriani}}, \citenamefont {{Barbarino}}, \citenamefont {{Bazilevskaya}}
  \emph {et~al.}}]{2013PhRvL.111h1102A}%
  \BibitemOpen
  \bibfield  {author} {\bibinfo {author} {\bibfnamefont {O.}~\bibnamefont
  {{Adriani}}}, \bibinfo {author} {\bibfnamefont {G.~C.}\ \bibnamefont
  {{Barbarino}}}, \bibinfo {author} {\bibfnamefont {G.~A.}\ \bibnamefont
  {{Bazilevskaya}}},  \emph {et~al.},\ }\href {\doibase
  10.1103/PhysRevLett.111.081102} {\bibfield  {journal} {\bibinfo  {journal}
  {Physical Review Letters}\ }\textbf {\bibinfo {volume} {111}},\ \bibinfo
  {eid} {081102} (\bibinfo {year} {2013})},\ \Eprint
  {http://arxiv.org/abs/1308.0133} {arXiv:1308.0133 [astro-ph.HE]} \BibitemShut
  {NoStop}%
\bibitem [{\citenamefont {{Accardo}}\ \emph {et~al.}(2014)\citenamefont
  {{Accardo}}, \citenamefont {{Aguilar}}, \citenamefont {{Aisa}}, \citenamefont
  {{Alvino}}, \citenamefont {{Ambrosi}}, \citenamefont {{Andeen}},
  \citenamefont {{Arruda}}, \citenamefont {{Attig}}, \citenamefont
  {{Azzarello}}, \citenamefont {{Bachlechner}} \emph {et~al.}}]{amsfrac}%
  \BibitemOpen
  \bibfield  {author} {\bibinfo {author} {\bibfnamefont {L.}~\bibnamefont
  {{Accardo}}}, \bibinfo {author} {\bibfnamefont {M.}~\bibnamefont
  {{Aguilar}}}, \bibinfo {author} {\bibfnamefont {D.}~\bibnamefont {{Aisa}}},
  \bibinfo {author} {\bibfnamefont {A.}~\bibnamefont {{Alvino}}}, \bibinfo
  {author} {\bibfnamefont {G.}~\bibnamefont {{Ambrosi}}}, \bibinfo {author}
  {\bibfnamefont {K.}~\bibnamefont {{Andeen}}}, \bibinfo {author}
  {\bibfnamefont {L.}~\bibnamefont {{Arruda}}}, \bibinfo {author}
  {\bibfnamefont {N.}~\bibnamefont {{Attig}}}, \bibinfo {author} {\bibfnamefont
  {P.}~\bibnamefont {{Azzarello}}}, \bibinfo {author} {\bibfnamefont
  {A.}~\bibnamefont {{Bachlechner}}},  \emph {et~al.},\ }\href {\doibase
  10.1103/PhysRevLett.113.121101} {\bibfield  {journal} {\bibinfo  {journal}
  {Physical Review Letters}\ }\textbf {\bibinfo {volume} {113}},\ \bibinfo
  {eid} {121101} (\bibinfo {year} {2014})}\BibitemShut {NoStop}%
\bibitem [{\citenamefont {{Delahaye}}\ \emph {et~al.}(2009)\citenamefont
  {{Delahaye}}, \citenamefont {{Lineros}}, \citenamefont {{Donato}},
  \citenamefont {{Fornengo}}, \citenamefont {{Lavalle}}, \citenamefont
  {{Salati}},\ and\ \citenamefont {{Taillet}}}]{dela09}%
  \BibitemOpen
  \bibfield  {author} {\bibinfo {author} {\bibfnamefont {T.}~\bibnamefont
  {{Delahaye}}}, \bibinfo {author} {\bibfnamefont {R.}~\bibnamefont
  {{Lineros}}}, \bibinfo {author} {\bibfnamefont {F.}~\bibnamefont {{Donato}}},
  \bibinfo {author} {\bibfnamefont {N.}~\bibnamefont {{Fornengo}}}, \bibinfo
  {author} {\bibfnamefont {J.}~\bibnamefont {{Lavalle}}}, \bibinfo {author}
  {\bibfnamefont {P.}~\bibnamefont {{Salati}}}, \ and\ \bibinfo {author}
  {\bibfnamefont {R.}~\bibnamefont {{Taillet}}},\ }\href {\doibase
  10.1051/0004-6361/200811130} {\bibfield  {journal} {\bibinfo  {journal}
  {\aap}\ }\textbf {\bibinfo {volume} {501}},\ \bibinfo {pages} {821} (\bibinfo
  {year} {2009})},\ \Eprint {http://arxiv.org/abs/0809.5268} {arXiv:0809.5268}
  \BibitemShut {NoStop}%
\bibitem [{\citenamefont {{Yuan}}\ \emph {et~al.}(2017)\citenamefont {{Yuan}},
  \citenamefont {{Lin}}, \citenamefont {{Fang}},\ and\ \citenamefont
  {{Bi}}}]{2017PhRvD..95h3007Y}%
  \BibitemOpen
  \bibfield  {author} {\bibinfo {author} {\bibfnamefont {Q.}~\bibnamefont
  {{Yuan}}}, \bibinfo {author} {\bibfnamefont {S.-J.}\ \bibnamefont {{Lin}}},
  \bibinfo {author} {\bibfnamefont {K.}~\bibnamefont {{Fang}}}, \ and\ \bibinfo
  {author} {\bibfnamefont {X.-J.}\ \bibnamefont {{Bi}}},\ }\href {\doibase
  10.1103/PhysRevD.95.083007} {\bibfield  {journal} {\bibinfo  {journal}
  {\prd}\ }\textbf {\bibinfo {volume} {95}},\ \bibinfo {eid} {083007} (\bibinfo
  {year} {2017})},\ \Eprint {http://arxiv.org/abs/1701.06149} {arXiv:1701.06149
  [astro-ph.HE]} \BibitemShut {NoStop}%
\bibitem [{\citenamefont {{Aguilar}}\ \emph {et~al.}(2016)\citenamefont
  {{Aguilar}}, \citenamefont {{Ali Cavasonza}}, \citenamefont {{Ambrosi}},
  \citenamefont {{Arruda}}, \citenamefont {{Attig}}, \citenamefont {{Aupetit}},
  \citenamefont {{Azzarello}}, \citenamefont {{Bachlechner}}, \citenamefont
  {{Barao}}, \citenamefont {{Barrau}},\ and\ \citenamefont
  {et~al.}}]{2016PhRvL.117w1102A}%
  \BibitemOpen
  \bibfield  {author} {\bibinfo {author} {\bibfnamefont {M.}~\bibnamefont
  {{Aguilar}}}, \bibinfo {author} {\bibfnamefont {L.}~\bibnamefont {{Ali
  Cavasonza}}}, \bibinfo {author} {\bibfnamefont {G.}~\bibnamefont
  {{Ambrosi}}}, \bibinfo {author} {\bibfnamefont {L.}~\bibnamefont {{Arruda}}},
  \bibinfo {author} {\bibfnamefont {N.}~\bibnamefont {{Attig}}}, \bibinfo
  {author} {\bibfnamefont {S.}~\bibnamefont {{Aupetit}}}, \bibinfo {author}
  {\bibfnamefont {P.}~\bibnamefont {{Azzarello}}}, \bibinfo {author}
  {\bibfnamefont {A.}~\bibnamefont {{Bachlechner}}}, \bibinfo {author}
  {\bibfnamefont {F.}~\bibnamefont {{Barao}}}, \bibinfo {author} {\bibfnamefont
  {A.}~\bibnamefont {{Barrau}}}, \ and\ \bibinfo {author} {\bibnamefont
  {et~al.}},\ }\href {\doibase 10.1103/PhysRevLett.117.231102} {\bibfield
  {journal} {\bibinfo  {journal} {Physical Review Letters}\ }\textbf {\bibinfo
  {volume} {117}},\ \bibinfo {eid} {231102} (\bibinfo {year}
  {2016})}\BibitemShut {NoStop}%
\bibitem [{\citenamefont {{Han}}\ and\ \citenamefont
  {{Qiao}}(1994)}]{1994A&A...288..759H}%
  \BibitemOpen
  \bibfield  {author} {\bibinfo {author} {\bibfnamefont {J.~L.}\ \bibnamefont
  {{Han}}}\ and\ \bibinfo {author} {\bibfnamefont {G.~J.}\ \bibnamefont
  {{Qiao}}},\ }\href@noop {} {\bibfield  {journal} {\bibinfo  {journal} {\aap}\
  }\textbf {\bibinfo {volume} {288}},\ \bibinfo {pages} {759} (\bibinfo {year}
  {1994})}\BibitemShut {NoStop}%
\bibitem [{\citenamefont {{Schlickeiser}}\ and\ \citenamefont
  {{Ruppel}}(2010)}]{schli10}%
  \BibitemOpen
  \bibfield  {author} {\bibinfo {author} {\bibfnamefont {R.}~\bibnamefont
  {{Schlickeiser}}}\ and\ \bibinfo {author} {\bibfnamefont {J.}~\bibnamefont
  {{Ruppel}}},\ }\href {\doibase 10.1088/1367-2630/12/3/033044} {\bibfield
  {journal} {\bibinfo  {journal} {New Journal of Physics}\ }\textbf {\bibinfo
  {volume} {12}},\ \bibinfo {eid} {033044} (\bibinfo {year} {2010})},\ \Eprint
  {http://arxiv.org/abs/0908.2183} {arXiv:0908.2183 [astro-ph.HE]} \BibitemShut
  {NoStop}%
\bibitem [{\citenamefont {{Ginzburg}}\ and\ \citenamefont
  {{Syrovatskii}}(1964)}]{1964ocr..book.....G}%
  \BibitemOpen
  \bibfield  {author} {\bibinfo {author} {\bibfnamefont {V.~L.}\ \bibnamefont
  {{Ginzburg}}}\ and\ \bibinfo {author} {\bibfnamefont {S.~I.}\ \bibnamefont
  {{Syrovatskii}}},\ }\href@noop {} {\emph {\bibinfo {title} {The Origin of
  Cosmic Rays, New York: Macmillan, 1964}}}\ (\bibinfo {year}
  {1964})\BibitemShut {NoStop}%
\bibitem [{\citenamefont {{Fang}}\ \emph
  {et~al.}(2017{\natexlab{a}})\citenamefont {{Fang}}, \citenamefont {{Wang}},
  \citenamefont {{Bi}}, \citenamefont {{Lin}},\ and\ \citenamefont
  {{Yin}}}]{2017ApJ...836..172F}%
  \BibitemOpen
  \bibfield  {author} {\bibinfo {author} {\bibfnamefont {K.}~\bibnamefont
  {{Fang}}}, \bibinfo {author} {\bibfnamefont {B.-B.}\ \bibnamefont {{Wang}}},
  \bibinfo {author} {\bibfnamefont {X.-J.}\ \bibnamefont {{Bi}}}, \bibinfo
  {author} {\bibfnamefont {S.-J.}\ \bibnamefont {{Lin}}}, \ and\ \bibinfo
  {author} {\bibfnamefont {P.-F.}\ \bibnamefont {{Yin}}},\ }\href {\doibase
  10.3847/1538-4357/aa5b93} {\bibfield  {journal} {\bibinfo  {journal} {\apj}\
  }\textbf {\bibinfo {volume} {836}},\ \bibinfo {eid} {172} (\bibinfo {year}
  {2017}{\natexlab{a}})},\ \Eprint {http://arxiv.org/abs/1611.10292}
  {arXiv:1611.10292 [astro-ph.HE]} \BibitemShut {NoStop}%
\bibitem [{\citenamefont {{Krymskii}}(1977)}]{1977DoSSR.234.1306K}%
  \BibitemOpen
  \bibfield  {author} {\bibinfo {author} {\bibfnamefont {G.~F.}\ \bibnamefont
  {{Krymskii}}},\ }\href@noop {} {\bibfield  {journal} {\bibinfo  {journal}
  {Akademiia Nauk SSSR Doklady}\ }\textbf {\bibinfo {volume} {234}},\ \bibinfo
  {pages} {1306} (\bibinfo {year} {1977})}\BibitemShut {NoStop}%
\bibitem [{\citenamefont {{Axford}}\ \emph {et~al.}(1977)\citenamefont
  {{Axford}}, \citenamefont {{Leer}},\ and\ \citenamefont
  {{Skadron}}}]{1977ICRC...11..132A}%
  \BibitemOpen
  \bibfield  {author} {\bibinfo {author} {\bibfnamefont {W.~I.}\ \bibnamefont
  {{Axford}}}, \bibinfo {author} {\bibfnamefont {E.}~\bibnamefont {{Leer}}}, \
  and\ \bibinfo {author} {\bibfnamefont {G.}~\bibnamefont {{Skadron}}},\
  }\href@noop {} {\bibfield  {journal} {\bibinfo  {journal} {International
  Cosmic Ray Conference}\ }\textbf {\bibinfo {volume} {11}},\ \bibinfo {pages}
  {132} (\bibinfo {year} {1977})}\BibitemShut {NoStop}%
\bibitem [{\citenamefont {{Bell}}(1978{\natexlab{a}})}]{1978MNRAS.182..147B}%
  \BibitemOpen
  \bibfield  {author} {\bibinfo {author} {\bibfnamefont {A.~R.}\ \bibnamefont
  {{Bell}}},\ }\href {\doibase 10.1093/mnras/182.2.147} {\bibfield  {journal}
  {\bibinfo  {journal} {\mnras}\ }\textbf {\bibinfo {volume} {182}},\ \bibinfo
  {pages} {147} (\bibinfo {year} {1978}{\natexlab{a}})}\BibitemShut {NoStop}%
\bibitem [{\citenamefont {{Bell}}(1978{\natexlab{b}})}]{1978MNRAS.182..443B}%
  \BibitemOpen
  \bibfield  {author} {\bibinfo {author} {\bibfnamefont {A.~R.}\ \bibnamefont
  {{Bell}}},\ }\href {\doibase 10.1093/mnras/182.3.443} {\bibfield  {journal}
  {\bibinfo  {journal} {\mnras}\ }\textbf {\bibinfo {volume} {182}},\ \bibinfo
  {pages} {443} (\bibinfo {year} {1978}{\natexlab{b}})}\BibitemShut {NoStop}%
\bibitem [{\citenamefont {{Blandford}}\ and\ \citenamefont
  {{Ostriker}}(1978)}]{1978ApJ...221L..29B}%
  \BibitemOpen
  \bibfield  {author} {\bibinfo {author} {\bibfnamefont {R.~D.}\ \bibnamefont
  {{Blandford}}}\ and\ \bibinfo {author} {\bibfnamefont {J.~P.}\ \bibnamefont
  {{Ostriker}}},\ }\href {\doibase 10.1086/182658} {\bibfield  {journal}
  {\bibinfo  {journal} {\apjl}\ }\textbf {\bibinfo {volume} {221}},\ \bibinfo
  {pages} {L29} (\bibinfo {year} {1978})}\BibitemShut {NoStop}%
\bibitem [{\citenamefont {{Caprioli}}\ \emph
  {et~al.}(2010{\natexlab{b}})\citenamefont {{Caprioli}}, \citenamefont
  {{Amato}},\ and\ \citenamefont {{Blasi}}}]{2010APh....33..307C}%
  \BibitemOpen
  \bibfield  {author} {\bibinfo {author} {\bibfnamefont {D.}~\bibnamefont
  {{Caprioli}}}, \bibinfo {author} {\bibfnamefont {E.}~\bibnamefont {{Amato}}},
  \ and\ \bibinfo {author} {\bibfnamefont {P.}~\bibnamefont {{Blasi}}},\ }\href
  {\doibase 10.1016/j.astropartphys.2010.03.001} {\bibfield  {journal}
  {\bibinfo  {journal} {Astroparticle Physics}\ }\textbf {\bibinfo {volume}
  {33}},\ \bibinfo {pages} {307} (\bibinfo {year} {2010}{\natexlab{b}})},\
  \Eprint {http://arxiv.org/abs/0912.2714} {arXiv:0912.2714 [astro-ph.HE]}
  \BibitemShut {NoStop}%
\bibitem [{\citenamefont {{Bell}}(2015)}]{2015MNRAS.447.2224B}%
  \BibitemOpen
  \bibfield  {author} {\bibinfo {author} {\bibfnamefont {A.~R.}\ \bibnamefont
  {{Bell}}},\ }\href {\doibase 10.1093/mnras/stu2596} {\bibfield  {journal}
  {\bibinfo  {journal} {\mnras}\ }\textbf {\bibinfo {volume} {447}},\ \bibinfo
  {pages} {2224} (\bibinfo {year} {2015})},\ \Eprint
  {http://arxiv.org/abs/1412.7294} {arXiv:1412.7294 [astro-ph.HE]} \BibitemShut
  {NoStop}%
\bibitem [{\citenamefont {{Zirakashvili}}\ and\ \citenamefont
  {{Aharonian}}(2007)}]{2007A&A...465..695Z}%
  \BibitemOpen
  \bibfield  {author} {\bibinfo {author} {\bibfnamefont {V.~N.}\ \bibnamefont
  {{Zirakashvili}}}\ and\ \bibinfo {author} {\bibfnamefont {F.}~\bibnamefont
  {{Aharonian}}},\ }\href {\doibase 10.1051/0004-6361:20066494} {\bibfield
  {journal} {\bibinfo  {journal} {\aap}\ }\textbf {\bibinfo {volume} {465}},\
  \bibinfo {pages} {695} (\bibinfo {year} {2007})},\ \Eprint
  {http://arxiv.org/abs/astro-ph/0612717} {astro-ph/0612717} \BibitemShut
  {NoStop}%
\bibitem [{\citenamefont {{Yamazaki}}\ \emph {et~al.}(2006)\citenamefont
  {{Yamazaki}}, \citenamefont {{Kohri}}, \citenamefont {{Bamba}}, \citenamefont
  {{Yoshida}}, \citenamefont {{Tsuribe}},\ and\ \citenamefont
  {{Takahara}}}]{yamazaki06}%
  \BibitemOpen
  \bibfield  {author} {\bibinfo {author} {\bibfnamefont {R.}~\bibnamefont
  {{Yamazaki}}}, \bibinfo {author} {\bibfnamefont {K.}~\bibnamefont {{Kohri}}},
  \bibinfo {author} {\bibfnamefont {A.}~\bibnamefont {{Bamba}}}, \bibinfo
  {author} {\bibfnamefont {T.}~\bibnamefont {{Yoshida}}}, \bibinfo {author}
  {\bibfnamefont {T.}~\bibnamefont {{Tsuribe}}}, \ and\ \bibinfo {author}
  {\bibfnamefont {F.}~\bibnamefont {{Takahara}}},\ }\href {\doibase
  10.1111/j.1365-2966.2006.10832.x} {\bibfield  {journal} {\bibinfo  {journal}
  {\mnras}\ }\textbf {\bibinfo {volume} {371}},\ \bibinfo {pages} {1975}
  (\bibinfo {year} {2006})},\ \Eprint {http://arxiv.org/abs/astro-ph/0601704}
  {astro-ph/0601704} \BibitemShut {NoStop}%
\bibitem [{\citenamefont {{Lorimer}}(2004)}]{l04}%
  \BibitemOpen
  \bibfield  {author} {\bibinfo {author} {\bibfnamefont {D.~R.}\ \bibnamefont
  {{Lorimer}}},\ }in\ \href@noop {} {\emph {\bibinfo {booktitle} {Young Neutron
  Stars and Their Environments}}},\ \bibinfo {series} {IAU Symposium}, Vol.\
  \bibinfo {volume} {218},\ \bibinfo {editor} {edited by\ \bibinfo {editor}
  {\bibfnamefont {F.}~\bibnamefont {{Camilo}}}\ and\ \bibinfo {editor}
  {\bibfnamefont {B.~M.}\ \bibnamefont {{Gaensler}}}}\ (\bibinfo {year}
  {2004})\ p.\ \bibinfo {pages} {105},\ \Eprint
  {http://arxiv.org/abs/astro-ph/0308501} {astro-ph/0308501} \BibitemShut
  {NoStop}%
\bibitem [{\citenamefont {{Green}}(2014)}]{green}%
  \BibitemOpen
  \bibfield  {author} {\bibinfo {author} {\bibfnamefont {D.~A.}\ \bibnamefont
  {{Green}}},\ }\href@noop {} {\bibfield  {journal} {\bibinfo  {journal}
  {Bulletin of the Astronomical Society of India}\ }\textbf {\bibinfo {volume}
  {42}},\ \bibinfo {pages} {47} (\bibinfo {year} {2014})},\ \Eprint
  {http://arxiv.org/abs/1409.0637} {arXiv:1409.0637 [astro-ph.HE]} \BibitemShut
  {NoStop}%
\bibitem [{\citenamefont {{Daugherty}}\ and\ \citenamefont
  {{Harding}}(1982)}]{1982ApJ...252..337D}%
  \BibitemOpen
  \bibfield  {author} {\bibinfo {author} {\bibfnamefont {J.~K.}\ \bibnamefont
  {{Daugherty}}}\ and\ \bibinfo {author} {\bibfnamefont {A.~K.}\ \bibnamefont
  {{Harding}}},\ }\href {\doibase 10.1086/159561} {\bibfield  {journal}
  {\bibinfo  {journal} {\apj}\ }\textbf {\bibinfo {volume} {252}},\ \bibinfo
  {pages} {337} (\bibinfo {year} {1982})}\BibitemShut {NoStop}%
\bibitem [{\citenamefont {{Crawford}}\ and\ \citenamefont
  {{Tiffany}}(2007)}]{2007AJ....134.1231C}%
  \BibitemOpen
  \bibfield  {author} {\bibinfo {author} {\bibfnamefont {F.}~\bibnamefont
  {{Crawford}}}\ and\ \bibinfo {author} {\bibfnamefont {C.~L.}\ \bibnamefont
  {{Tiffany}}},\ }\href {\doibase 10.1086/521020} {\bibfield  {journal}
  {\bibinfo  {journal} {\aj}\ }\textbf {\bibinfo {volume} {134}},\ \bibinfo
  {pages} {1231} (\bibinfo {year} {2007})},\ \Eprint
  {http://arxiv.org/abs/0706.3182} {arXiv:0706.3182} \BibitemShut {NoStop}%
\bibitem [{\citenamefont {{Hou}}\ \emph {et~al.}(2011)\citenamefont {{Hou}},
  \citenamefont {{Dumora}}, \citenamefont {{Lemoine-Goumard}}, \citenamefont
  {{Grondin}}, \citenamefont {{Smith}}, \citenamefont {{for the Fermi Large
  Area Telescope Collaboration}},\ and\ \citenamefont {{Fermi Pulsar Timing
  Consortium}}}]{2011arXiv1110.1210H}%
  \BibitemOpen
  \bibfield  {author} {\bibinfo {author} {\bibfnamefont {X.}~\bibnamefont
  {{Hou}}}, \bibinfo {author} {\bibfnamefont {D.}~\bibnamefont {{Dumora}}},
  \bibinfo {author} {\bibfnamefont {M.}~\bibnamefont {{Lemoine-Goumard}}},
  \bibinfo {author} {\bibfnamefont {M.~.}\ \bibnamefont {{Grondin}}}, \bibinfo
  {author} {\bibfnamefont {D.~A.}\ \bibnamefont {{Smith}}}, \bibinfo {author}
  {\bibnamefont {{for the Fermi Large Area Telescope Collaboration}}}, \ and\
  \bibinfo {author} {\bibfnamefont {f.~t.}\ \bibnamefont {{Fermi Pulsar Timing
  Consortium}}},\ }\href@noop {} {\bibfield  {journal} {\bibinfo  {journal}
  {ArXiv e-prints}\ } (\bibinfo {year} {2011})},\ \Eprint
  {http://arxiv.org/abs/1110.1210} {arXiv:1110.1210 [astro-ph.HE]} \BibitemShut
  {NoStop}%
\bibitem [{\citenamefont {{Y{\"u}ksel}}\ \emph {et~al.}(2009)\citenamefont
  {{Y{\"u}ksel}}, \citenamefont {{Kistler}},\ and\ \citenamefont
  {{Stanev}}}]{2009PhRvL.103e1101Y}%
  \BibitemOpen
  \bibfield  {author} {\bibinfo {author} {\bibfnamefont {H.}~\bibnamefont
  {{Y{\"u}ksel}}}, \bibinfo {author} {\bibfnamefont {M.~D.}\ \bibnamefont
  {{Kistler}}}, \ and\ \bibinfo {author} {\bibfnamefont {T.}~\bibnamefont
  {{Stanev}}},\ }\href {\doibase 10.1103/PhysRevLett.103.051101} {\bibfield
  {journal} {\bibinfo  {journal} {Physical Review Letters}\ }\textbf {\bibinfo
  {volume} {103}},\ \bibinfo {eid} {051101} (\bibinfo {year} {2009})},\ \Eprint
  {http://arxiv.org/abs/0810.2784} {arXiv:0810.2784} \BibitemShut {NoStop}%
\bibitem [{\citenamefont {{Linden}}\ and\ \citenamefont
  {{Profumo}}(2013)}]{2013ApJ...772...18L}%
  \BibitemOpen
  \bibfield  {author} {\bibinfo {author} {\bibfnamefont {T.}~\bibnamefont
  {{Linden}}}\ and\ \bibinfo {author} {\bibfnamefont {S.}~\bibnamefont
  {{Profumo}}},\ }\href {\doibase 10.1088/0004-637X/772/1/18} {\bibfield
  {journal} {\bibinfo  {journal} {\apj}\ }\textbf {\bibinfo {volume} {772}},\
  \bibinfo {eid} {18} (\bibinfo {year} {2013})},\ \Eprint
  {http://arxiv.org/abs/1304.1791} {arXiv:1304.1791 [astro-ph.HE]} \BibitemShut
  {NoStop}%
\bibitem [{\citenamefont {{Hooper}}\ \emph {et~al.}(2017)\citenamefont
  {{Hooper}}, \citenamefont {{Cholis}}, \citenamefont {{Linden}},\ and\
  \citenamefont {{Fang}}}]{2017arXiv170208436H}%
  \BibitemOpen
  \bibfield  {author} {\bibinfo {author} {\bibfnamefont {D.}~\bibnamefont
  {{Hooper}}}, \bibinfo {author} {\bibfnamefont {I.}~\bibnamefont {{Cholis}}},
  \bibinfo {author} {\bibfnamefont {T.}~\bibnamefont {{Linden}}}, \ and\
  \bibinfo {author} {\bibfnamefont {K.}~\bibnamefont {{Fang}}},\ }\href@noop {}
  {\bibfield  {journal} {\bibinfo  {journal} {ArXiv e-prints}\ } (\bibinfo
  {year} {2017})},\ \Eprint {http://arxiv.org/abs/1702.08436} {arXiv:1702.08436
  [astro-ph.HE]} \BibitemShut {NoStop}%
\bibitem [{\citenamefont {{Abdo}}\ \emph {et~al.}(2009)\citenamefont {{Abdo}},
  \citenamefont {{Allen}}, \citenamefont {{Aune}} \emph
  {et~al.}}]{2009ApJ...700L.127A}%
  \BibitemOpen
  \bibfield  {author} {\bibinfo {author} {\bibfnamefont {A.~A.}\ \bibnamefont
  {{Abdo}}}, \bibinfo {author} {\bibfnamefont {B.~T.}\ \bibnamefont {{Allen}}},
  \bibinfo {author} {\bibfnamefont {T.}~\bibnamefont {{Aune}}},  \emph
  {et~al.},\ }\href {\doibase 10.1088/0004-637X/700/2/L127} {\bibfield
  {journal} {\bibinfo  {journal} {\apjl}\ }\textbf {\bibinfo {volume} {700}},\
  \bibinfo {pages} {L127} (\bibinfo {year} {2009})},\ \Eprint
  {http://arxiv.org/abs/0904.1018} {arXiv:0904.1018 [astro-ph.HE]} \BibitemShut
  {NoStop}%
\bibitem [{\citenamefont {{Abeysekara}}\ \emph {et~al.}(2017)\citenamefont
  {{Abeysekara}}, \citenamefont {{Albert}}, \citenamefont {{Alfaro}} \emph
  {et~al.}}]{2017ApJ...843...40A}%
  \BibitemOpen
  \bibfield  {author} {\bibinfo {author} {\bibfnamefont {A.~U.}\ \bibnamefont
  {{Abeysekara}}}, \bibinfo {author} {\bibfnamefont {A.}~\bibnamefont
  {{Albert}}}, \bibinfo {author} {\bibfnamefont {R.}~\bibnamefont {{Alfaro}}},
  \emph {et~al.},\ }\href {\doibase 10.3847/1538-4357/aa7556} {\bibfield
  {journal} {\bibinfo  {journal} {\apj}\ }\textbf {\bibinfo {volume} {843}},\
  \bibinfo {eid} {40} (\bibinfo {year} {2017})},\ \Eprint
  {http://arxiv.org/abs/1702.02992} {arXiv:1702.02992 [astro-ph.HE]}
  \BibitemShut {NoStop}%
\bibitem [{\citenamefont {Abeysekara}\ \emph {et~al.}(2017)\citenamefont
  {Abeysekara}, \citenamefont {Albert}, \citenamefont {Alfaro} \emph
  {et~al.}}]{Abeysekara911}%
  \BibitemOpen
  \bibfield  {author} {\bibinfo {author} {\bibfnamefont {A.~U.}\ \bibnamefont
  {Abeysekara}}, \bibinfo {author} {\bibfnamefont {A.}~\bibnamefont {Albert}},
  \bibinfo {author} {\bibfnamefont {R.}~\bibnamefont {Alfaro}},  \emph
  {et~al.},\ }\href {\doibase 10.1126/science.aan4880} {\bibfield  {journal}
  {\bibinfo  {journal} {Science}\ }\textbf {\bibinfo {volume} {358}},\ \bibinfo
  {pages} {911} (\bibinfo {year} {2017})},\ \Eprint
  {http://arxiv.org/abs/1711.06223} {arXiv:1711.06223} \BibitemShut {NoStop}%
\bibitem [{\citenamefont {{Hooper}}\ and\ \citenamefont
  {{Linden}}(2017)}]{2017arXiv171107482H}%
  \BibitemOpen
  \bibfield  {author} {\bibinfo {author} {\bibfnamefont {D.}~\bibnamefont
  {{Hooper}}}\ and\ \bibinfo {author} {\bibfnamefont {T.}~\bibnamefont
  {{Linden}}},\ }\href@noop {} {\bibfield  {journal} {\bibinfo  {journal}
  {ArXiv e-prints}\ } (\bibinfo {year} {2017})},\ \Eprint
  {http://arxiv.org/abs/1711.07482} {arXiv:1711.07482 [astro-ph.HE]}
  \BibitemShut {NoStop}%
\bibitem [{\citenamefont {{Kothes}}\ \emph {et~al.}(2008)\citenamefont
  {{Kothes}}, \citenamefont {{Landecker}}, \citenamefont {{Reich}},
  \citenamefont {{Safi-Harb}},\ and\ \citenamefont
  {{Arzoumanian}}}]{2008ApJ...687..516K}%
  \BibitemOpen
  \bibfield  {author} {\bibinfo {author} {\bibfnamefont {R.}~\bibnamefont
  {{Kothes}}}, \bibinfo {author} {\bibfnamefont {T.~L.}\ \bibnamefont
  {{Landecker}}}, \bibinfo {author} {\bibfnamefont {W.}~\bibnamefont
  {{Reich}}}, \bibinfo {author} {\bibfnamefont {S.}~\bibnamefont
  {{Safi-Harb}}}, \ and\ \bibinfo {author} {\bibfnamefont {Z.}~\bibnamefont
  {{Arzoumanian}}},\ }\href {\doibase 10.1086/591653} {\bibfield  {journal}
  {\bibinfo  {journal} {\apj}\ }\textbf {\bibinfo {volume} {687}},\ \bibinfo
  {eid} {516-531} (\bibinfo {year} {2008})},\ \Eprint
  {http://arxiv.org/abs/0807.0811} {arXiv:0807.0811} \BibitemShut {NoStop}%
\bibitem [{\citenamefont {{Reynolds}}\ \emph {et~al.}(2017)\citenamefont
  {{Reynolds}}, \citenamefont {{Pavlov}}, \citenamefont {{Kargaltsev}},
  \citenamefont {{Klingler}}, \citenamefont {{Renaud}},\ and\ \citenamefont
  {{Mereghetti}}}]{2017SSRv..207..175R}%
  \BibitemOpen
  \bibfield  {author} {\bibinfo {author} {\bibfnamefont {S.~P.}\ \bibnamefont
  {{Reynolds}}}, \bibinfo {author} {\bibfnamefont {G.~G.}\ \bibnamefont
  {{Pavlov}}}, \bibinfo {author} {\bibfnamefont {O.}~\bibnamefont
  {{Kargaltsev}}}, \bibinfo {author} {\bibfnamefont {N.}~\bibnamefont
  {{Klingler}}}, \bibinfo {author} {\bibfnamefont {M.}~\bibnamefont
  {{Renaud}}}, \ and\ \bibinfo {author} {\bibfnamefont {S.}~\bibnamefont
  {{Mereghetti}}},\ }\href {\doibase 10.1007/s11214-017-0356-6} {\bibfield
  {journal} {\bibinfo  {journal} {\ssr}\ }\textbf {\bibinfo {volume} {207}},\
  \bibinfo {pages} {175} (\bibinfo {year} {2017})},\ \Eprint
  {http://arxiv.org/abs/1705.08897} {arXiv:1705.08897 [astro-ph.HE]}
  \BibitemShut {NoStop}%
\bibitem [{\citenamefont {{Plucinsky}}(2009)}]{plucinsky09}%
  \BibitemOpen
  \bibfield  {author} {\bibinfo {author} {\bibfnamefont {P.~P.}\ \bibnamefont
  {{Plucinsky}}},\ }in\ \href {\doibase 10.1063/1.3211819} {\emph {\bibinfo
  {booktitle} {American Institute of Physics Conference Series}}},\ Vol.\
  \bibinfo {volume} {1156},\ \bibinfo {editor} {edited by\ \bibinfo {editor}
  {\bibfnamefont {R.~K.}\ \bibnamefont {{Smith}}}, \bibinfo {editor}
  {\bibfnamefont {S.~L.}\ \bibnamefont {{Snowden}}}, \ and\ \bibinfo {editor}
  {\bibfnamefont {K.~D.}\ \bibnamefont {{Kuntz}}}}\ (\bibinfo {year} {2009})\
  pp.\ \bibinfo {pages} {231--235}\BibitemShut {NoStop}%
\bibitem [{\citenamefont {{Taylor}}\ \emph {et~al.}(1993)\citenamefont
  {{Taylor}}, \citenamefont {{Manchester}},\ and\ \citenamefont
  {{Lyne}}}]{atnf}%
  \BibitemOpen
  \bibfield  {author} {\bibinfo {author} {\bibfnamefont {J.~H.}\ \bibnamefont
  {{Taylor}}}, \bibinfo {author} {\bibfnamefont {R.~N.}\ \bibnamefont
  {{Manchester}}}, \ and\ \bibinfo {author} {\bibfnamefont {A.~G.}\
  \bibnamefont {{Lyne}}},\ }\href {\doibase 10.1086/191832} {\bibfield
  {journal} {\bibinfo  {journal} {\apjs}\ }\textbf {\bibinfo {volume} {88}},\
  \bibinfo {pages} {529} (\bibinfo {year} {1993})}\BibitemShut {NoStop}%
\bibitem [{\citenamefont {{Abdollahi}}\ \emph {et~al.}(2017)\citenamefont
  {{Abdollahi}}, \citenamefont {{Ackermann}}, \citenamefont {{Ajello}} \emph
  {et~al.}}]{2017PhRvL.118i1103A}%
  \BibitemOpen
  \bibfield  {author} {\bibinfo {author} {\bibfnamefont {S.}~\bibnamefont
  {{Abdollahi}}}, \bibinfo {author} {\bibfnamefont {M.}~\bibnamefont
  {{Ackermann}}}, \bibinfo {author} {\bibfnamefont {M.}~\bibnamefont
  {{Ajello}}},  \emph {et~al.},\ }\href {\doibase
  10.1103/PhysRevLett.118.091103} {\bibfield  {journal} {\bibinfo  {journal}
  {\prl}\ }\textbf {\bibinfo {volume} {118}},\ \bibinfo {eid} {091103}
  (\bibinfo {year} {2017})},\ \Eprint {http://arxiv.org/abs/1703.01073}
  {arXiv:1703.01073 [astro-ph.HE]} \BibitemShut {NoStop}%
\bibitem [{\citenamefont {{Atoyan}}\ \emph {et~al.}(1995)\citenamefont
  {{Atoyan}}, \citenamefont {{Aharonian}},\ and\ \citenamefont
  {{V{\"o}lk}}}]{1995PhRvD..52.3265A}%
  \BibitemOpen
  \bibfield  {author} {\bibinfo {author} {\bibfnamefont {A.~M.}\ \bibnamefont
  {{Atoyan}}}, \bibinfo {author} {\bibfnamefont {F.~A.}\ \bibnamefont
  {{Aharonian}}}, \ and\ \bibinfo {author} {\bibfnamefont {H.~J.}\ \bibnamefont
  {{V{\"o}lk}}},\ }\href {\doibase 10.1103/PhysRevD.52.3265} {\bibfield
  {journal} {\bibinfo  {journal} {\prd}\ }\textbf {\bibinfo {volume} {52}},\
  \bibinfo {pages} {3265} (\bibinfo {year} {1995})}\BibitemShut {NoStop}%
\bibitem [{\citenamefont {{Shen}}\ and\ \citenamefont
  {{Mao}}(1971)}]{1971ApL.....9..169S}%
  \BibitemOpen
  \bibfield  {author} {\bibinfo {author} {\bibfnamefont {C.~S.}\ \bibnamefont
  {{Shen}}}\ and\ \bibinfo {author} {\bibfnamefont {C.~Y.}\ \bibnamefont
  {{Mao}}},\ }\href@noop {} {\bibfield  {journal} {\bibinfo  {journal}
  {Astrophysical Letters}\ }\textbf {\bibinfo {volume} {9}},\ \bibinfo {pages}
  {169} (\bibinfo {year} {1971})}\BibitemShut {NoStop}%
\bibitem [{\citenamefont {{Manconi}}\ \emph {et~al.}(2017)\citenamefont
  {{Manconi}}, \citenamefont {{Di Mauro}},\ and\ \citenamefont
  {{Donato}}}]{2017JCAP...01..006M}%
  \BibitemOpen
  \bibfield  {author} {\bibinfo {author} {\bibfnamefont {S.}~\bibnamefont
  {{Manconi}}}, \bibinfo {author} {\bibfnamefont {M.}~\bibnamefont {{Di
  Mauro}}}, \ and\ \bibinfo {author} {\bibfnamefont {F.}~\bibnamefont
  {{Donato}}},\ }\href {\doibase 10.1088/1475-7516/2017/01/006} {\bibfield
  {journal} {\bibinfo  {journal} {\jcap}\ }\textbf {\bibinfo {volume} {1}},\
  \bibinfo {eid} {006} (\bibinfo {year} {2017})},\ \Eprint
  {http://arxiv.org/abs/1611.06237} {arXiv:1611.06237 [astro-ph.HE]}
  \BibitemShut {NoStop}%
\bibitem [{\citenamefont {{Fang}}\ \emph
  {et~al.}(2017{\natexlab{b}})\citenamefont {{Fang}}, \citenamefont {{Bi}},\
  and\ \citenamefont {{Yin}}}]{2017arXiv170603745F}%
  \BibitemOpen
  \bibfield  {author} {\bibinfo {author} {\bibfnamefont {K.}~\bibnamefont
  {{Fang}}}, \bibinfo {author} {\bibfnamefont {X.-J.}\ \bibnamefont {{Bi}}}, \
  and\ \bibinfo {author} {\bibfnamefont {P.-f.}\ \bibnamefont {{Yin}}},\
  }\href@noop {} {\bibfield  {journal} {\bibinfo  {journal} {ArXiv e-prints}\ }
  (\bibinfo {year} {2017}{\natexlab{b}})},\ \Eprint
  {http://arxiv.org/abs/1706.03745} {arXiv:1706.03745 [astro-ph.HE]}
  \BibitemShut {NoStop}%
\bibitem [{\citenamefont {{Zhang}}\ \emph {et~al.}(2014)\citenamefont
  {{Zhang}}, \citenamefont {{Adriani}}, \citenamefont {{Albergo}} \emph
  {et~al.}}]{2014SPIE.9144E..0XZ}%
  \BibitemOpen
  \bibfield  {author} {\bibinfo {author} {\bibfnamefont {S.~N.}\ \bibnamefont
  {{Zhang}}}, \bibinfo {author} {\bibfnamefont {O.}~\bibnamefont {{Adriani}}},
  \bibinfo {author} {\bibfnamefont {S.}~\bibnamefont {{Albergo}}},  \emph
  {et~al.},\ }in\ \href {\doibase 10.1117/12.2055280} {\emph {\bibinfo
  {booktitle} {Space Telescopes and Instrumentation 2014: Ultraviolet to Gamma
  Ray}}},\ \bibinfo {series} {Proceedings of the SPIE}, Vol.\ \bibinfo {volume}
  {9144}\ (\bibinfo {year} {2014})\ p.\ \bibinfo {pages} {91440X},\ \Eprint
  {http://arxiv.org/abs/1407.4866} {arXiv:1407.4866 [astro-ph.IM]} \BibitemShut
  {NoStop}%
\bibitem [{\citenamefont {{Gaensler}}\ and\ \citenamefont
  {{Slane}}(2006)}]{pwn06}%
  \BibitemOpen
  \bibfield  {author} {\bibinfo {author} {\bibfnamefont {B.~M.}\ \bibnamefont
  {{Gaensler}}}\ and\ \bibinfo {author} {\bibfnamefont {P.~O.}\ \bibnamefont
  {{Slane}}},\ }\href {\doibase 10.1146/annurev.astro.44.051905.092528}
  {\bibfield  {journal} {\bibinfo  {journal} {\araa}\ }\textbf {\bibinfo
  {volume} {44}},\ \bibinfo {pages} {17} (\bibinfo {year} {2006})},\ \Eprint
  {http://arxiv.org/abs/astro-ph/0601081} {astro-ph/0601081} \BibitemShut
  {NoStop}%
\bibitem [{\citenamefont {{van der Swaluw}}\ \emph {et~al.}(2004)\citenamefont
  {{van der Swaluw}}, \citenamefont {{Downes}},\ and\ \citenamefont
  {{Keegan}}}]{2004A&A...420..937V}%
  \BibitemOpen
  \bibfield  {author} {\bibinfo {author} {\bibfnamefont {E.}~\bibnamefont {{van
  der Swaluw}}}, \bibinfo {author} {\bibfnamefont {T.~P.}\ \bibnamefont
  {{Downes}}}, \ and\ \bibinfo {author} {\bibfnamefont {R.}~\bibnamefont
  {{Keegan}}},\ }\href {\doibase 10.1051/0004-6361:20035700} {\bibfield
  {journal} {\bibinfo  {journal} {\aap}\ }\textbf {\bibinfo {volume} {420}},\
  \bibinfo {pages} {937} (\bibinfo {year} {2004})},\ \Eprint
  {http://arxiv.org/abs/astro-ph/0311388} {astro-ph/0311388} \BibitemShut
  {NoStop}%
\bibitem [{\citenamefont {{Malyshev}}\ \emph {et~al.}(2009)\citenamefont
  {{Malyshev}}, \citenamefont {{Cholis}},\ and\ \citenamefont
  {{Gelfand}}}]{maly09}%
  \BibitemOpen
  \bibfield  {author} {\bibinfo {author} {\bibfnamefont {D.}~\bibnamefont
  {{Malyshev}}}, \bibinfo {author} {\bibfnamefont {I.}~\bibnamefont
  {{Cholis}}}, \ and\ \bibinfo {author} {\bibfnamefont {J.}~\bibnamefont
  {{Gelfand}}},\ }\href {\doibase 10.1103/PhysRevD.80.063005} {\bibfield
  {journal} {\bibinfo  {journal} {\prd}\ }\textbf {\bibinfo {volume} {80}},\
  \bibinfo {eid} {063005} (\bibinfo {year} {2009})},\ \Eprint
  {http://arxiv.org/abs/0903.1310} {arXiv:0903.1310 [astro-ph.HE]} \BibitemShut
  {NoStop}%
\bibitem [{\citenamefont {{Pacini}}\ and\ \citenamefont
  {{Salvati}}(1973)}]{1973ApJ...186..249P}%
  \BibitemOpen
  \bibfield  {author} {\bibinfo {author} {\bibfnamefont {F.}~\bibnamefont
  {{Pacini}}}\ and\ \bibinfo {author} {\bibfnamefont {M.}~\bibnamefont
  {{Salvati}}},\ }\href {\doibase 10.1086/152495} {\bibfield  {journal}
  {\bibinfo  {journal} {\apj}\ }\textbf {\bibinfo {volume} {186}},\ \bibinfo
  {pages} {249} (\bibinfo {year} {1973})}\BibitemShut {NoStop}%
\bibitem [{\citenamefont {{Manchester}}\ \emph {et~al.}(2005)\citenamefont
  {{Manchester}}, \citenamefont {{Hobbs}}, \citenamefont {{Teoh}},\ and\
  \citenamefont {{Hobbs}}}]{2005AJ....129.1993M}%
  \BibitemOpen
  \bibfield  {author} {\bibinfo {author} {\bibfnamefont {R.~N.}\ \bibnamefont
  {{Manchester}}}, \bibinfo {author} {\bibfnamefont {G.~B.}\ \bibnamefont
  {{Hobbs}}}, \bibinfo {author} {\bibfnamefont {A.}~\bibnamefont {{Teoh}}}, \
  and\ \bibinfo {author} {\bibfnamefont {M.}~\bibnamefont {{Hobbs}}},\ }\href
  {\doibase 10.1086/428488} {\bibfield  {journal} {\bibinfo  {journal} {\aj}\
  }\textbf {\bibinfo {volume} {129}},\ \bibinfo {pages} {1993} (\bibinfo {year}
  {2005})},\ \Eprint {http://arxiv.org/abs/astro-ph/0412641} {astro-ph/0412641}
  \BibitemShut {NoStop}%
\bibitem [{\citenamefont {{Blondin}}\ \emph {et~al.}(2001)\citenamefont
  {{Blondin}}, \citenamefont {{Chevalier}},\ and\ \citenamefont
  {{Frierson}}}]{blondin01}%
  \BibitemOpen
  \bibfield  {author} {\bibinfo {author} {\bibfnamefont {J.~M.}\ \bibnamefont
  {{Blondin}}}, \bibinfo {author} {\bibfnamefont {R.~A.}\ \bibnamefont
  {{Chevalier}}}, \ and\ \bibinfo {author} {\bibfnamefont {D.~M.}\ \bibnamefont
  {{Frierson}}},\ }\href {\doibase 10.1086/324042} {\bibfield  {journal}
  {\bibinfo  {journal} {\apj}\ }\textbf {\bibinfo {volume} {563}},\ \bibinfo
  {pages} {806} (\bibinfo {year} {2001})},\ \Eprint
  {http://arxiv.org/abs/astro-ph/0107076} {astro-ph/0107076} \BibitemShut
  {NoStop}%
\bibitem [{\citenamefont {{Hinton}}\ \emph {et~al.}(2011)\citenamefont
  {{Hinton}}, \citenamefont {{Funk}}, \citenamefont {{Parsons}},\ and\
  \citenamefont {{Ohm}}}]{hinton11}%
  \BibitemOpen
  \bibfield  {author} {\bibinfo {author} {\bibfnamefont {J.~A.}\ \bibnamefont
  {{Hinton}}}, \bibinfo {author} {\bibfnamefont {S.}~\bibnamefont {{Funk}}},
  \bibinfo {author} {\bibfnamefont {R.~D.}\ \bibnamefont {{Parsons}}}, \ and\
  \bibinfo {author} {\bibfnamefont {S.}~\bibnamefont {{Ohm}}},\ }\href
  {\doibase 10.1088/2041-8205/743/1/L7} {\bibfield  {journal} {\bibinfo
  {journal} {\apjl}\ }\textbf {\bibinfo {volume} {743}},\ \bibinfo {eid} {L7}
  (\bibinfo {year} {2011})},\ \Eprint {http://arxiv.org/abs/1111.2036}
  {arXiv:1111.2036 [astro-ph.HE]} \BibitemShut {NoStop}%
\end{thebibliography}%

\end{document}